\documentclass[10pt,journal,epsfig]{IEEEtran}

\usepackage[dvips]{graphicx}
\usepackage{graphicx}
\usepackage{amssymb}
\usepackage{cite}
\usepackage{subfigure}
\usepackage{amsmath}

\begin{document}

\title{Two-Dimensional Pattern-Coupled Sparse Bayesian Learning via Generalized Approximate Message Passing}

\author{Jun Fang, Lizao Zhang, and Hongbin Li,~\IEEEmembership{Senior Member,~IEEE}
\thanks{Jun Fang, and Lizao Zhang are with the National Key Laboratory
of Science and Technology on Communications, University of
Electronic Science and Technology of China, Chengdu 611731, China,
Email: JunFang@uestc.edu.cn}
\thanks{Hongbin Li is
with the Department of Electrical and Computer Engineering,
Stevens Institute of Technology, Hoboken, NJ 07030, USA, E-mail:
Hongbin.Li@stevens.edu}
\thanks{This work was supported in part by the National Science
Foundation of China under Grant 61172114, and the National Science
Foundation under Grant ECCS-1408182.}}

\maketitle

\begin{abstract}
We consider the problem of recovering two-dimensional (2-D)
block-sparse signals with \emph{unknown} cluster patterns.
Two-dimensional block-sparse patterns arise naturally in many
practical applications such as foreground detection and inverse
synthetic aperture radar imaging. To exploit the block-sparse
structure, we introduce a 2-D pattern-coupled hierarchical
Gaussian prior model to characterize the statistical pattern
dependencies among neighboring coefficients. Unlike the
conventional hierarchical Gaussian prior model where each
coefficient is associated independently with a unique
hyperparameter, the pattern-coupled prior for each coefficient not
only involves its own hyperparameter, but also its immediate
neighboring hyperparameters. Thus the sparsity patterns of
neighboring coefficients are related to each other and the
hierarchical model has the potential to encourage 2-D
structured-sparse solutions. An expectation-maximization (EM)
strategy is employed to obtain the maximum a posterior (MAP)
estimate of the hyperparameters, along with the posterior
distribution of the sparse signal. In addition, the generalized
approximate message passing (GAMP) algorithm is embedded into the
EM framework to efficiently compute an approximation of the
posterior distribution of hidden variables, which results in a
significant reduction in computational complexity. Numerical
results are provided to illustrate the effectiveness of the
proposed algorithm.
\end{abstract}


\section{Introduction}
Compressed sensing is a recently emerged technique for signal
sampling and data acquisition which enables to recover sparse
signals from undersampled linear measurements
\begin{align}
\boldsymbol{y}=\boldsymbol{Ax}+\boldsymbol{w} \label{data-model}
\end{align}
where $\boldsymbol{A}\in\mathbb{R}^{M\times N}$ is a sampling
matrix with $M\ll N$, $\boldsymbol{x}$ denotes an $N$-dimensional
sparse signal, and $\boldsymbol{w}$ denotes the additive noise.
The problem has been extensively studied and a variety of
algorithms, e.g. the orthogonal matching pursuit (OMP) algorithm
\cite{TroppGilbert07}, the basis pursuit (BP) method
\cite{ChenDonoho98}, the iterative reweighted $\ell_1$ and
$\ell_2$ algorithms \cite{WipfNagaranjan10}, and the sparse
Bayesian learning method \cite{Tipping01,JiXue08,YangXie13} were
proposed. In many practical applications, in addition to the
sparse structure, sparse signals may exhibit two-dimensional
cluster patterns that can be utilized to enhance the recovery
performance. For example, the target of interest in the synthetic
aperture radar/inverse synthetic aperture radar (SAR/ISAR) images
often demonstrates continuity in both the range and cross-range
domains \cite{WangZhao14}. In video surveillance, the foreground
image exhibits a cluster pattern since the foreground objects
(humans, cars, text etc.) generally occupy a small continuous
region of the scene \cite{Cevher11}. Besides these, block-sparsity
is also present in temporal observations of a time-varying
block-sparse signal whose support varies slowly over time
\cite{VaswaniLu10}.




Analyses \cite{EldarMishali09,BaraniukCevher10,EldarKuppinger10}
show that exploiting the inherent block-sparse structure not only
leads to relaxed conditions for exact reconstruction, but also
helps improve the recovery performance considerably. A number of
algorithms have been proposed for recovering block-sparse signals
over the past few years, e.g., block-OMP \cite{EldarKuppinger10},
mixed $\ell_2/\ell_1$ norm-minimization \cite{EldarMishali09},
group LASSO \cite{YuanLin06}, model-based CoSaMP
\cite{BaraniukCevher10}, and block-sparse Bayesian learning
\cite{WipfRao07,ZhangRao11}. These algorithms, however, require
\emph{a priori} knowledge of the block partition (e.g. the number
of blocks and location of each block) such that the coefficients
in each block are grouped together and enforced to share a common
sparsity pattern. In practice, the prior information about the
block partition of sparse signals is often unavailable, especially
for two-dimensional signals since the block partition of a
two-dimensional signal involves not only the location but also the
shape of each block. For example, foreground images have irregular
and unpredictable cluster patterns which are very difficult to be
estimated \emph{a priori}. To address this difficulty, a few
sophisticated Bayesian methods which do not need the knowledge of
the block partition were developed. In \cite{HeCarin09}, a
``spike-and-slab'' prior model was proposed, where by introducing
dependencies among mixing weights, the prior model has the
potential to encourage sparsity and promote a tree structure
simultaneously. This ``spike-and-slab'' prior model was later
extended to accommodate block-sparse signals
\cite{YuSun12,WangZhao14}. Nevertheless, for the
``spike-and-slab'' prior introduced in \cite{HeCarin09,YuSun12},
the posterior distribution cannot be derived analytically, and a
Markov chain Monte Carlo (MCMC) sampling method has to be employed
for Bayesian inference. In \cite{PelegEldar12,DremeauHerzet12}, a
graphical prior, also referred to as the ``Boltzmann machine'', is
employed as a prior on the sparsity support in order to induce
statistical dependencies between atoms. With such a prior, the
maximum a posterior (MAP) estimator requires an exhaustive search
over all possible sparsity patterns. To overcome the
intractability of the combinatorial search, a greedy method
\cite{PelegEldar12} and a variational mean-field approximation
method \cite{DremeauHerzet12} were developed to approximate the
MAP. In \cite{ZhangRao13}, to cope with the unknown cluster
pattern, an expanded model is employed by assuming that the
original sparse signal is a superposition of a number of
overlapping blocks, and the coefficients in each block share the
same sparsity pattern. Conventional block sparse Bayesian learning
algorithms such as those in \cite{ZhangRao11} can then be applied
to the expanded model.

Recently in \cite{FangShen15}, we proposed a pattern-coupled
hierarchical Gaussian prior model to exploit the unknown
block-sparse structure. Unlike the conventional hierarchical
Gaussian prior model \cite{Tipping01,JiXue08,YangXie13} where each
coefficient is associated independently with a unique
hyperparameter, the pattern-coupled prior for each coefficient not
only involves its own hyperparameter, but also its immediate
neighboring hyperparameters. This pattern-coupled hierarchical
model is effective and flexible to capture any underlying
block-sparse structures, without requiring the prior knowledge of
the block partition. Numerical results show that the
pattern-coupled sparse Bayesian learning (PC-SBL) method renders
competitive performance for block-sparse signal recovery.
Nevertheless, a major drawback of the method is that it requires
computing an $N\times N$ matrix inverse at each iteration, and
thus has a cubic complexity in terms of the signal dimension. This
high computational cost prohibits its application to problems with
even moderate dimensions. Also, \cite{FangShen15} only considers
recovery of one-dimensional block-sparse signals. In this paper,
we generalize the pattern-coupled hierarchical model to the
two-dimensional (2-D) scenario in order to leverage block-sparse
patterns arising from 2-D sparse signals. To address the
computational issue, we resort to the generalized approximate
message passing (GAMP) technique \cite{Rangan11} and develop a
computationally efficient method. Specifically, the algorithm is
developed within an expectation-maximization (EM) framework, using
the GAMP to efficiently compute an approximation of the posterior
distribution of hidden variables. The hyperparameters associated
with the hierarchical Gaussian prior are learned by iteratively
maximizing the Q-function which is calculated based on the
posterior approximation obtained from the GAMP. Simulation results
show that the proposed method presents superior recovery
performance for block-sparse signals, meanwhile achieving a
significant reduction in computational complexity.

The rest of the paper is organized as follows. In Section
\ref{sec:model}, we introduce a 2-D pattern coupled hierarchical
Gaussian framework to model the sparse prior and the pattern
dependencies among the neighboring coefficients. In Section
\ref{sec:algorithm}, a GAMP-based EM algorithm is developed to
obtain the maximum a posterior (MAP) estimate of the
hyperparameters, along with the posterior distribution of the
sparse signal. Simulation results are provided in Section
\ref{sec:simulation}, followed by concluding remarks in Section
\ref{sec:conclusions}.


\section{Bayesian Model}
\label{sec:model} We consider the problem of recovering a
two-dimensional block-sparse signal
$\boldsymbol{X}\in\mathbb{R}^{Q\times L}$ from compressed noisy
measurements
\begin{align}
\boldsymbol{y}=\boldsymbol{f}(\boldsymbol{X})+\boldsymbol{w}
\label{data-model}
\end{align}
where $\boldsymbol{y}\in\mathbb{R}^M$ denotes the compressed
measurement vector, $\boldsymbol{f}(\cdot)$ is a linear map:
$\mathbb{R}^{Q\times L}\rightarrow\mathbb{R}^M$, with $M\ll
N\triangleq QL$, and $\boldsymbol{w}\in\mathbb{R}^M$ is an
additive multivariate Gaussian noise with zero mean and covariance
matrix $\sigma^{2}\boldsymbol{I}$. Let
$\boldsymbol{x}\triangleq\text{vec}(\boldsymbol{X})$, the linear
map $\boldsymbol{f}(\boldsymbol{X})$ can generally be expressed as
\begin{align}
\boldsymbol{f}(\boldsymbol{X})=\boldsymbol{A}\boldsymbol{x}
\label{linear-map}
\end{align}
where $\boldsymbol{A}\in\mathbb{R}^{M\times N}$ denotes the
measurement matrix. In the special case where
$\boldsymbol{f}(\boldsymbol{X})=\text{vec}(\boldsymbol{BX})$, then
we have
$\boldsymbol{A}\triangleq\boldsymbol{I}\otimes\boldsymbol{B}$, in
which $\otimes$ stands for the Kronecker product. The above model
(\ref{data-model}) arises in image applications where signals are
multi-dimensional in nature, or in the scenario where multiple
snapshots of a time-varying sparse signal are available. In these
applications, signals usually exhibit two-dimensional cluster
patterns that can be utilized to improve the recovery accuracy. To
leverage the underlying block-sparse structures, we introduce a
2-D pattern-coupled Gaussian prior model which is a generalization
of our previous work \cite{FangShen15}. Before proceeding, we
provide a brief review of the conventional hierarchical Gaussian
prior model \cite{Tipping01}, and some of its extensions.


\subsection{Review of Conventional Gaussian Prior Model}
For ease of exposition, we consider the prior model for the
two-dimensional signal $\boldsymbol{X}$ instead of its
one-dimensional form $\boldsymbol{x}$. Let $x_{q,l}$ denote the
$(q,l)$th entry of $\boldsymbol{X}$. In the conventional sparse
Bayesian learning framework \cite{Tipping01}, a two-layer
hierarchical Gaussian prior was employed to promote the sparsity
of the solution. In the first layer, coefficients $\{x_{q,l}\}$ of
$\boldsymbol{X}$ are assigned a Gaussian prior distribution
\begin{align}
p(\boldsymbol{X}|\boldsymbol{\alpha})
=\prod_{q=1}^Q\prod_{l=1}^L\mathcal{N}(x_{q,l}|0,\alpha_{q,l}^{-1})
\label{hm-1}
\end{align}
where $\alpha_{q,l}$ is a non-negative hyperparameter controlling
the sparsity of the coefficient $x_{q,l}$. The second layer
specifies Gamma distributions as hyperpriors over the
hyperparameters $\boldsymbol{\alpha}\triangleq\{\alpha_{q,l}\}$,
i.e.
\begin{align}
p(\boldsymbol{\alpha})=\prod_{q=1}^Q\prod_{l=1}^L\text{Gamma}(\alpha_{q,l}|a,b)
\label{hm-2}
\end{align}
As discussed in \cite{Tipping01}, for properly chosen $a$ and $b$,
this hyperprior allows the posterior mean of $\alpha_{q,l}$ to
become arbitrarily large. As a consequence, the associated
coefficient $x_{q,l}$ will be driven to zero, thus yielding a
sparse solution. This conventional hierarchical model, however,
does not encourage structured-sparse solutions since the sparsity
of each coefficient is determined by its own hyperparameter and
the hyperparameters are independent of each other. In
\cite{WipfRao07,ZhangRao11}, the above hierarchical model was
generalized to deal with block-sparse signals, in which a group of
coefficients sharing the same sparsity pattern are assigned a
multivariate Gaussian prior parameterized by a common
hyperparameter. Nevertheless, this model requires the knowledge of
the block partition to determine which coefficients should be
grouped and assigned a common hyperparameter.


\subsection{Proposed 2-D Pattern-Coupled Hierarchical Model}
To exploit the 2-D block-sparse structure, we utilize the fact
that the sparsity patterns of neighboring coefficients are
statistically dependent. To capture the pattern dependencies among
neighboring coefficients, the Gaussian prior for each coefficient
$x_{q,l}$ not only involves its own hyperparameter $\alpha_{q,l}$,
but also its immediate neighbor hyperparameters. Specifically, a
prior over $\boldsymbol{X}$ is given by
\begin{align}
p(\boldsymbol{X}|\boldsymbol{\alpha})=
\prod_{q=1}^Q\prod_{l=1}^L\mathcal{N}(x_{q,l}|0,\delta_{q,l}^{-1})
\label{pc-model}
\end{align}
where
\begin{align}
\delta_{q,l}\triangleq\alpha_{q,l}+\beta\sum_{(i,j)\in N_{(q,l)}}
\alpha_{i,j} \label{delta-mn}
\end{align}
in which $N_{(q,l)}$ denotes the neighborhood of the grid point
$(q,l)$, i.e.
$N_{(q,l)}\triangleq\{(q,l-1),(q,l+1),(q-1,l),(q+1,l)\}$\footnote{Note
that for edge grid points, they only have two or three immediate
neighboring points, in which case the definition of $N_{(q,l)}$
changes accordingly.}, and $\beta\in [0,1]$ is a parameter
indicating the pattern relevance between the coefficient $x_{q,l}$
and its neighboring coefficients. Clearly, this model is an
extension of our previous prior model \cite{FangShen15} to the
two-dimensional case. When $\beta=0$, the prior model
(\ref{pc-model}) reduces to the conventional sparse Bayesian
learning model. When $\beta>0$, we see that the sparsity of each
coefficient $x_{q,l}$ is not only controlled by the hyperparameter
$\alpha_{q,l}$, but also by the neighboring hyperparameters
$S_{\alpha_{q,l}}\triangleq\{\alpha_{i,j}|(i,j)\in N_{(q,l)}\}$.
The coefficient $x_{q,l}$ will be driven to zero if $\alpha_{q,l}$
or any of its neighboring hyperparameters goes to infinity. In
other words, suppose $\alpha_{q,l}$ approaches infinity, then not
only its corresponding coefficient $x_{q,l}$ will be driven to
zero, the neighboring coefficients
$S_{x_{q,l}}\triangleq\{x_{i,j}|(i,j)\in N_{(q,l)}\}$ will
decrease to zero as well. We see that the sparsity patterns of
neighboring coefficients are related to each other through their
shared hyperparameters. On the other hand, for any pair of
neighboring coefficients, each of them has its own hyperparameters
that are not shared by the other coefficient. Hence, no
coefficients are pre-specified to share a common sparsity pattern,
which enables the prior to provide flexibility to model any
block-sparse structures.

Following \cite{Tipping01}, we use Gamma distributions as
hyperpriors over the hyperparameters $\{\alpha_{q,l}\}$, i.e.
\begin{align}
p(\boldsymbol{\alpha})=\prod_{q=1}^Q\prod_{l=1}^L\text{Gamma}(\alpha_{q,l}|a,b)
\label{alpha-prior}
\end{align}
where we set $a>1$, and $b=10^{-6}$. The choice of $a$ will be
elaborated later in our paper. Also, the noise variance
$\sigma^2\triangleq 1/\gamma$ is assumed unknown, and to estimate
this parameter, we place a Gamma hyperprior over $\gamma$, i.e.
\begin{align}
p(\gamma)=\text{Gamma}(\gamma|c,d) \label{gamma-prior}
\end{align}
where we set $c=1$ and $d=10^{-6}$.





\section{Proposed Algorithm} \label{sec:algorithm}
We now proceed to perform Bayesian inference for the proposed
pattern-coupled hierarchical model. The following model is
considered since the linear map $\boldsymbol{f}(\boldsymbol{X})$
can be expressed as
$\boldsymbol{f}(\boldsymbol{X})=\boldsymbol{A}\boldsymbol{x}$
\begin{align}
\boldsymbol{y}=\boldsymbol{A}\boldsymbol{x}+\boldsymbol{w}
\end{align}
We first translate the prior for the two-dimensional signal
$\boldsymbol{X}$ to a prior for its one-dimensional form
$\boldsymbol{x}$. From (\ref{pc-model}), the prior over
$\boldsymbol{x}$ can be expressed as
\begin{align}
p(\boldsymbol{x}|\boldsymbol{\alpha})=
\prod_{n=1}^N\mathcal{N}(x_{n}|0,\eta_{n}^{-1}) \label{x-prior}
\end{align}
where
\begin{align}
\eta_n\triangleq\alpha_n+\beta\sum_{i\in N_{(n)}}\alpha_i
\label{eta-definition}
\end{align}
in which $N_{(n)}$ denotes the neighbors of the point $(q,l)$ on
the two-dimensional grid, i.e. $N_{(n)}\triangleq \{(l-2)Q+q,
lQ+q, (l-1)Q+q-1, (l-1)Q+q+1\}$\footnote{For edge grid points
which have only two or three neighboring points, the definition of
$N_{(n)}$ changes accordingly.}. The relation between $n$ and
$(q,l)$ is given by $n=(l-1)Q+q$, that is, $l=\lceil n/Q\rceil$,
and $q=n\phantom{0}\text{mod}\phantom{0}Q$, in which $\lceil
\cdot\rceil$ denotes the ceiling operator. Note that for
notational convenience, we, with a slight abuse of notation, use
$x_n$ to denote the $n$th entry of $\boldsymbol{x}$ and $\alpha_n$
to denote the hyperparameter associated with the coefficient
$x_n$. Also, let $\boldsymbol{\alpha}\triangleq\{\alpha_n\}$ since
its exact meaning remains unaltered. From (\ref{alpha-prior}), we
have
\begin{align}
p(\boldsymbol{\alpha})=\prod_{n=1}^N\text{Gamma}(\alpha_{n}|a,b)
\label{alpha-prior-new}
\end{align}

An expectation-maximization (EM) algorithm can be developed for
learning the sparse signal $\boldsymbol{x}$ as well as the
hyperparameters $\{\boldsymbol{\alpha},\gamma\}$. In the EM
formulation, the signal $\boldsymbol{x}$ is treated as hidden
variables, and we iteratively maximize a lower bound on the
posterior probability
$p(\boldsymbol{\alpha},\gamma|\boldsymbol{y})$ (this lower bound
is also referred to as the Q-function). Briefly speaking, the
algorithm alternates between an E-step and a M-step. In the
E-step, we need to compute the posterior distribution of
$\boldsymbol{x}$ conditioned on the observed data and the
hyperparameters estimated from the $t$th iteration, i.e.
\begin{align}
p(\boldsymbol{x}|\boldsymbol{y},\boldsymbol{\alpha}^{(t)},\gamma^{(t)})\propto
p(\boldsymbol{x}|\boldsymbol{\alpha}^{(t)})p(\boldsymbol{y}|\boldsymbol{x},\gamma^{(t)})
\end{align}
It can be readily verified that the posterior
$p(\boldsymbol{x}|\boldsymbol{y},\boldsymbol{\alpha}^{(t)},\gamma^{(t)})$
follows a Gaussian distribution with its mean and covariance
matrix given respectively by
\begin{align}
\boldsymbol{\mu}=&\gamma^{(t)}\boldsymbol{\Phi}\boldsymbol{A}^T\boldsymbol{y}
\\
\boldsymbol{\Phi}=&
(\gamma^{(t)}\boldsymbol{A}^T\boldsymbol{A}+\boldsymbol{D})^{-1}
\label{x-posterior}
\end{align}
where
$\boldsymbol{D}\triangleq\text{diag}(\eta_1^{(t)},\ldots,\eta_N^{(t)})$.
The Q-function, i.e.
$E_{\boldsymbol{x}|\boldsymbol{y},\boldsymbol{\alpha}^{(t)},\gamma^{(t)}}[\log
p(\boldsymbol{\alpha},\gamma|\boldsymbol{y})]$, can then be
computed, where the operator
$E_{\boldsymbol{x}|\boldsymbol{y},\boldsymbol{\alpha}^{(t)},\gamma^{(t)}}[\cdot]$
denotes the expectation with respect to the posterior distribution
$p(\boldsymbol{x}|\boldsymbol{y},\boldsymbol{\alpha}^{(t)},\gamma^{(t)})$.
In the M-step, we maximize the Q-function with respect to the
hyperparameters $\{\boldsymbol{\alpha},\gamma\}$.

It can be seen that the EM algorithm, at each iteration, requires
to update the posterior distribution
$p(\boldsymbol{x}|\boldsymbol{y},\boldsymbol{\alpha}^{(t)},\gamma^{(t)})$,
which involves computing an $N\times N$ matrix inverse. Thus the
EM-based algorithm has a computational complexity of
$\mathcal{O}(N^3)$ flops, and therefore is not suitable for many
real-world applications involving large dimensions. In the
following, we will develop a computationally efficient algorithm
by resorting to the generalized approximate message passing (GAMP)
technique \cite{Rangan11}. GAMP is a very-low-complexity Bayesian
iterative technique recently developed \cite{Rangan11} for
obtaining an approximation of the posterior distribution
$p(\boldsymbol{x}|\boldsymbol{y},\boldsymbol{\alpha}^{(t)},\gamma^{(t)})$.
It therefore can naturally be embedded within the EM framework to
replace the computation of the true posterior distribution. From
GAMP's point of view, the hyperparameters
$\{\boldsymbol{\alpha},\gamma\}$ are considered as known. The
hyperparameters can be updated in the M-step based on the
approximate posterior distribution of $\boldsymbol{x}$. We now
proceed to derive the GAMP algorithm for the pattern-coupled
Gaussian hierarchical prior model.


\subsection{Pattern-Coupled Hierarchical Gaussian GAMP}
GAMP was developed in a message passing-based framework. By using
central-limit-theorem approximations, message passing between
variable nodes and factor nodes can be greatly simplified, and the
loopy belief propagation on the underlying factor graph can be
efficiently performed. As noted in \cite{DonohoMaleki10,Rangan11},
the central-limit-theorem approximations become exact in the
large-system limit under i.i.d. zero-mean sub-Gaussian
$\boldsymbol{A}$.


For notational convenience, let
$\boldsymbol{\theta}\triangleq\{\boldsymbol{\alpha},\gamma\}$
denote the hyperparameters. Firstly, GAMP assumes posterior
independence among hidden variables $\{x_n\}$ and approximates the
true posterior distribution
$p(x_n|\boldsymbol{y},\boldsymbol{\theta})$ by
\begin{align}
\hat{p}(x_n|\boldsymbol{y},\hat{r}_n,\tau_n^r,\boldsymbol{\theta})
=&\frac{p(x_n|\boldsymbol{\theta})\mathcal{N}(x_n|\hat{r}_n,\tau_n^r)}{\int_x
p(x_n|\boldsymbol{\theta})\mathcal{N}(x_n|\hat{r}_n,\tau_n^r)}
\label{eqn-1}
\end{align}
where $\hat{r}_n$ and $\tau_n^r$ are quantities iteratively
updated during the iterative process of the GAMP algorithm. Here,
we have dropped their explicit dependence on the iteration number
$k$ for simplicity. Substituting (\ref{x-prior}) into
(\ref{eqn-1}), it can be easily verified that the approximate
posterior
$\hat{p}(x_n|\boldsymbol{y},\hat{r}_n,\tau_n^r,\boldsymbol{\theta})$
follows a Gaussian distribution with its mean and variance given
respectively as
\begin{align}
\mu_n^x&\triangleq\frac{\hat{r}_n}{1+\eta_n\tau_n^r} \label{x-post-mean} \\
\phi_n^x&\triangleq\frac{\tau_n^r}{1+\eta_n\tau_n^r}
\label{x-post-var}
\end{align}
Another approximation is made to the noiseless output
$z_m\triangleq\boldsymbol{a}_m^T\boldsymbol{x}$, where
$\boldsymbol{a}_m^T$ denotes the $m$th row of $\boldsymbol{A}$.
GAMP approximates the true marginal posterior
$p(z_m|\boldsymbol{y},\boldsymbol{\theta})$ by
\begin{align}
\hat{p}(z_m|\boldsymbol{y},\hat{p}_m,\tau_m^p,\boldsymbol{\theta})=\frac{p(y_m|z_m,\boldsymbol{\theta})
\mathcal{N}(z_m|\hat{p}_m,\tau_m^p)}{\int_z
p(y_m|z_m,\boldsymbol{\theta})
\mathcal{N}(z_m|\hat{p}_m,\tau_m^p)}
\end{align}
where $\hat{p}_m$ and $\tau_m^p$ are quantities iteratively
updated during the iterative process of the GAMP algorithm. Again,
here we dropped their explicit dependence on the iteration number
$k$. Under the additive white Gaussian noise assumption, we have
$p(y_m|z_m,\boldsymbol{\theta})=\mathcal{N}(y_m|z_m,1/\gamma)$.
Thus
$\hat{p}(z_m|\boldsymbol{y},\hat{p}_m,\tau_m^p,\boldsymbol{\theta})$
also follows a Gaussian distribution with its mean and variance
given by
\begin{align}
\mu_m^z\triangleq&\frac{\tau_m^p\gamma
y_m+\hat{p}_m}{1+\gamma\tau_m^p} \label{z-post-mean}
\\
\phi_m^z\triangleq&\frac{\tau_m^p}{1+\gamma\tau_m^p}
\label{z-post-var}
\end{align}

With the above approximations, we can now define the following two
scalar functions: $g_{\text{in}}(\cdot)$ and
$g_{\text{out}}(\cdot)$ that are used in the GAMP algorithm. The
input scalar function $g_{\text{in}}(\cdot)$ is simply defined as
the posterior mean $\mu_n^x$ \cite{Rangan11}, i.e.
\begin{align}
g_{\text{in}}(\hat{r}_n,\tau_n^r,
\boldsymbol{\theta})=\mu_n^x=\frac{\hat{r}_n}{1+\eta_n\tau_n^r}
\end{align}
The scaled partial derivative of $\tau_n^r
g_{\text{in}}(\hat{r}_n,\tau_n^r,\boldsymbol{\theta})$ with
respect to $\hat{r}_n$ is the posterior variance $\phi_n^x$, i.e.
\begin{align}
\tau_n^r\frac{\partial}{\partial\hat{r}_n}g_{\text{in}}(\hat{r}_n,\tau_n^r,\boldsymbol{\theta})
=\phi_n^x=\frac{\tau_n^r}{1+\eta_n\tau_n^r}
\end{align}
The output scalar function $g_{\text{out}}(\cdot)$ is related to
the posterior mean $\mu_m^z$ as follows
\begin{align}
g_{\text{out}}(\hat{p}_m,\tau_m^p,\boldsymbol{\theta})=\frac{1}{\tau_m^p}(\mu_m^z-\hat{p}_m)=
\frac{1}{\tau_m^p}\bigg(\frac{\tau_m^p\gamma
y_m+\hat{p}_m}{1+\gamma\tau_m^p}-\hat{p}_m\bigg)
\end{align}
The partial derivative of
$g_{\text{out}}(\hat{p}_m,\tau_m^p,\boldsymbol{\theta})$ is
related to the posterior variance $\phi_m^z$ in the following way
\begin{align}
\tau_m^p\frac{\partial}{\partial\hat{p}_m}g_{\text{out}}(\hat{p}_m,\tau_m^p,\boldsymbol{\theta})=
\frac{\phi_m^z-\tau_m^p}{\tau_m^p}=\frac{-\gamma\tau_m^p}{1+\gamma\tau_m^p}
\end{align}
Given the above definitions of $g_{\text{in}}(\cdot)$ and
$g_{\text{out}}(\cdot)$, the GAMP algorithm tailored to the
considered sparse signal estimation problem can now be summarized
as follows (details of the derivation of the GAMP algorithm can be
found in \cite{Rangan11}), in which $a_{mn}$ denotes the $(m,n)$th
entry of $\boldsymbol{A}$, $\mu_n^x(k)$ and $\phi_n^{x}(k)$ denote
the posterior mean and variance of $x_n$ at iteration $k$,
respectively.

\begin{center}

\vspace{0cm} \noindent
\begin{tabular}{p{8cm}}
\hline
\textbf{GAMP Algorithm} \\
\hline Initialization: given $\boldsymbol{\theta}^{(t)}$; set
$k=0$, $\hat{s}_m^{(-1)}=0,\forall m\in\{1,\ldots,M\}$;
$\{\mu_n^x(k)\}_{n=1}^N$ and $\{\phi_n^{x}(k)\}_{n=1}^N$ are
initialized as the mean and variance of the prior distribution. \\
Repeat the following steps until
$\sum_{n}|\mu_n^x(k+1)-\mu_n^x(k)|^2\leq\epsilon$, where
$\epsilon$ is a pre-specified error tolerance. \\
Step 1. $\forall m\in\{1,\ldots,M\}$: \\
$\begin{aligned}
\qquad\qquad\quad \hat{z}_m(k)=&\sum_n a_{mn}\mu_n^x(k)\\
\tau_m^p(k)=&\sum_n a_{mn}^2\phi_n^{x}(k) \\
\hat{p}_m(k)=&\hat{z}_m(k)-\tau_m^p(k)\hat{s}_m(k-1)\\
\end{aligned}$ \\
Step 2. $\forall m\in\{1,\ldots,M\}$: \\
$\begin{aligned} \qquad\qquad\quad
\hat{s}_m(k)=&g_{\text{out}}(\hat{p}_m(k),\tau^p_m(k),\boldsymbol{\theta}^{(t)})\\
\tau^s_m(k)=&-\frac{\partial}{\partial\hat{p}_m}g_{\text{out}}(\hat{p}_m(k),\tau^p_m(k),\boldsymbol{\theta}^{(t)})\\
\end{aligned}$ \\
Step 3. $\forall n\in\{1,\ldots,N\}$: \\
$\begin{aligned} \qquad\qquad\quad
\tau_n^r(k)=&\left(\sum_{m}a_{mn}^2\tau^s_m(k)\right)^{-1} \\
\hat{r}_n(k)=&\mu_n^x(k)+\tau_n^r(k)\sum_{m}a_{mn}\hat{s}_m(k) \\
\end{aligned}$ \\
Step 4. $\forall n\in\{1,\ldots,N\}$: \\
$\begin{aligned} \qquad\qquad\quad
\mu_n^x(k+1)=&g_{\text{in}}(\hat{r}_n(k),\tau_n^r(k),\boldsymbol{\theta}^{(t)})
\\
\phi_n^{x}(k+1)=&\tau_n^r(k)\frac{\partial}{\partial\hat{r}_n}g_{\text{in}}(\hat{r}_n(k),\tau_n^r(k),\boldsymbol{\theta}^{(t)})
\\
\end{aligned}$ \\
Output: $\{\hat{r}_n(k_0),\tau_n^r(k_0)\}$,
$\{\hat{p}_m(k_0),\tau_m^p(k_0)\}$, and
$\{\mu_n^x(k_0+1),\phi_n^{x}(k_0+1)\}$, where $k_0$ stands for the
last iteration.\\
\hline
\end{tabular}
\vspace{0.3cm}
\end{center}

We have now derived an efficient algorithm to obtain approximate
posterior distributions for the variables $\boldsymbol{x}$ and
$\boldsymbol{z}\triangleq\boldsymbol{Ax}$. We see that the GAMP
algorithm no longer needs to compute the aforementioned matrix
inverse. The dominating operations in each iteration is the simple
matrix-vector multiplications, which scale as $\mathcal{O}(MN)$.
Thus the computational complexity is significantly reduced. In the
following, we discuss how to update the hyperparameters via the
EM.

\subsection{Hyperparameter Learning via EM}
Given the current estimates of the hyperparameters
$\boldsymbol{\theta}^{(t)}\triangleq\{\boldsymbol{\alpha}^{(t)},\gamma^{(t)}\}$
and the observed data $\boldsymbol{y}$, the E-step computes the
expected value (with respect to the hidden variables
$\boldsymbol{x}$) of the complete log-posterior of
$\{\boldsymbol{\alpha},\gamma\}$, i.e.
$E_{\boldsymbol{x}|\boldsymbol{y},\boldsymbol{\alpha}^{(t)},\gamma^{(t)}}[\log
p(\boldsymbol{\alpha},\gamma|\boldsymbol{x},\boldsymbol{y})]$,
where the operator
$E_{\boldsymbol{x}|\boldsymbol{y},\boldsymbol{\alpha}^{(t)},\gamma^{(t)}}[\cdot]$
denotes the expectation with respect to the posterior distribution
$p(\boldsymbol{x}|\boldsymbol{y},\boldsymbol{\alpha}^{(t)},\gamma^{(t)})$.
This complete log-posterior is also referred to as the Q-function.
The hyperparameters $\{\boldsymbol{\alpha},\gamma\}$ are then
estimated by maximizing the Q-function, i.e.
\begin{align}
\{\boldsymbol{\alpha}^{(t+1)},\gamma^{(t+1)}\}=&\arg\max_{\boldsymbol{\alpha},\gamma}
Q(\boldsymbol{\alpha},\gamma|\boldsymbol{\alpha}^{(t)},\gamma^{(t)})
\nonumber\\
=&\arg\max_{\boldsymbol{\alpha},\gamma}
E_{\boldsymbol{x}|\boldsymbol{y},\boldsymbol{\alpha}^{(t)},\gamma^{(t)}}[\log
p(\boldsymbol{\alpha},\gamma|\boldsymbol{x},\boldsymbol{y})]
\end{align}
Since
\begin{align}
p(\boldsymbol{\alpha},\gamma|\boldsymbol{x},\boldsymbol{y})\propto
p(\boldsymbol{\alpha})p(\boldsymbol{x}|\boldsymbol{\alpha})p(\gamma)p(\boldsymbol{y}|\boldsymbol{x},\gamma)
\end{align}
the Q-function can be decomposed into a summation of two terms
\begin{align}
Q(\boldsymbol{\alpha},\gamma|\boldsymbol{\alpha}^{(t)},\gamma^{(t)})=&
E_{\boldsymbol{x}|\boldsymbol{y},\boldsymbol{\alpha}^{(t)},\gamma^{(t)}}[\log
p(\boldsymbol{\alpha})p(\boldsymbol{x}|\boldsymbol{\alpha})]
\nonumber\\
&+
E_{\boldsymbol{x}|\boldsymbol{y},\boldsymbol{\alpha}^{(t)},\gamma^{(t)}}[\log
p(\gamma)p(\boldsymbol{y}|\boldsymbol{x},\gamma)] \nonumber\\
\triangleq
&Q(\boldsymbol{\alpha}|\boldsymbol{\alpha}^{(t)},\gamma^{(t)})+Q(\gamma|\boldsymbol{\alpha}^{(t)},\gamma^{(t)})
\label{Q-function}
\end{align}
We observe that in the above Q-function, the hyperparameters
$\boldsymbol{\alpha}$ and $\gamma$ are separated from each other.
This decouples the estimation of $\boldsymbol{\alpha}$ and
$\gamma$ into two independent optimization problems. We first
examine the update of $\boldsymbol{\alpha}$, i.e.
\begin{align}
\boldsymbol{\alpha}^{(t+1)} =
\arg\max_{\boldsymbol{\alpha}}Q(\boldsymbol{\alpha}|\boldsymbol{\alpha}^{(t)},\gamma^{(t)})
\label{opt-1}
\end{align}
Recalling (\ref{x-prior})--(\ref{alpha-prior-new}),
$Q(\boldsymbol{\alpha}|\boldsymbol{\alpha}^{(t)},\gamma^{(t)})$
can be expressed as
\begin{align}
&Q(\boldsymbol{\alpha}|\boldsymbol{\alpha}^{(t)},\gamma^{(t)})\nonumber\\
\propto&
\sum_{n=1}^N\left((a-1)\log\alpha_{n}-b\alpha_{n}+\frac{1}{2}\log\eta_{n}
-\frac{1}{2}\eta_{n} \langle x_{n}^2\rangle\right)
\label{Q-function-alpha}
\end{align}
where $\langle x_{n}^2\rangle$ denotes the expectation with
respect to the posterior distribution
$p(\boldsymbol{x}|\boldsymbol{y},\boldsymbol{\alpha}^{(t)},\gamma^{(t)})$.
Here we use
$\hat{p}(x_n|\boldsymbol{y},\hat{r}_n(k_0),\tau_n^r(k_0),\boldsymbol{\theta}^{(t)})$,
i.e. the approximate posterior distribution of $x_n$ obtained from
the GAMP algorithm, to replace the true posterior distribution
$p(\boldsymbol{x}|\boldsymbol{y},\boldsymbol{\alpha}^{(t)},\gamma^{(t)})$
in computing the expectation. Since
$\hat{p}(x_n|\boldsymbol{y},\hat{r}_n(k_0),\tau_n^r(k_0),\boldsymbol{\theta}^{(t)})$
follows a Gaussian distribution with its mean and variance given
by (\ref{x-post-mean})--(\ref{x-post-var}), we have
\begin{align}
\langle x_n^2\rangle=
\frac{(\hat{r}_n(k_0))^2}{(1+\eta_n^{(t)}\tau_n^r(k_0))^2}+\frac{\tau_n^r(k_0)}{1+\eta_n^{(t)}\tau_n^r(k_0)}
\end{align}

We see that in the Q-function (\ref{Q-function-alpha}),
hyperparameters are entangled with each other due to the logarithm
term $\log\eta_{n}$ (note that $\eta_n$, defined in
(\ref{eta-definition}), is a function of $\boldsymbol{\alpha}$).
In this case, an analytical solution to the optimization
(\ref{opt-1}) is difficult to obtain. Gradient descent methods can
certainly be used to search for the optimal solution.
Nevertheless, for gradient descent methods, there is no explicit
formula for the hyperparameter update. Also, gradient-based
methods involve higher computational complexity as compared with
an analytical update rule. Here we consider an alternative
strategy which aims at finding a simple, analytical sub-optimal
solution of (\ref{opt-1}). Such an analytical sub-optimal solution
can be obtained by examining the optimality condition of
(\ref{opt-1}). Suppose
$\boldsymbol{\alpha}^{\ast}=\{\alpha_{n}^{\ast}\}$ is the optimal
solution of (\ref{opt-1}). Then the first derivative of the
Q-function with respect to $\boldsymbol{\alpha}$ equals to zero at
the optimal point, i.e.
\begin{align}
\frac{\partial Q(\boldsymbol{\alpha}|\boldsymbol{\theta}^{(t)})}
{\partial
{{\alpha_n}}}|_{\boldsymbol\alpha=\boldsymbol{\alpha^{\ast}}}
=\frac{a-1}{\alpha_n^{\ast}}-b+\frac{1}{2}\nu_n^{\ast}-\frac{1}{2}\omega_n=0
\label{eqn-2}
\end{align}
where
\begin{align}
\nu_{n}^{\ast}\triangleq &
\frac{1}{\eta_{n}^{\ast}}+\beta\sum_{i\in
N_{(n)}}\frac{1}{\eta_{i}^{\ast}}
\\
\omega_{n}\triangleq &\langle x_n^2\rangle+\beta\sum_{i\in
N_{(n)}}\langle x_i^2\rangle
\end{align}
Since all hyperparameters $\{\alpha_n\}$ and $\beta$ are
non-negative, it can be easily verified
$(1/\alpha_{n}^{\ast})>(1/\eta_{n}^{\ast})>0$, and
$(1/\beta\alpha_{n}^{\ast})>(1/\eta_{i}^{\ast})>0$ for $i\in
N_{(n)}$. Therefore $\nu_{n}^{\ast}$ is bounded by
\begin{align}
\frac{5}{\alpha_{n}^{\ast}}>\nu_{n}^{\ast}>0
\end{align}
Consequently we have
\begin{align}
\frac{a+1.5}{\alpha_{n}^{\ast}}>\frac{a-1}{\alpha_{n}^{\ast}}+\frac{1}{2}\nu_{n}^{\ast}>\frac{a-1}{\alpha_{n}^{\ast}}
\label{eqn-3}
\end{align}
Combining (\ref{eqn-2}) and (\ref{eqn-3}), we reach that
$\alpha_{n}^{\ast}$ is within the range
\begin{align}
\alpha_{n}^{\ast}\in
\left[\frac{a-1}{0.5\omega_{n}+b},\frac{a+1.5}{0.5\omega_{n}+b}\right]
\quad \forall n
\end{align}
A sub-optimal solution to (\ref{opt-1}) can therefore simply be
chosen as
\begin{align}
\alpha_{n}^{(t+1)}=\frac{a-1}{0.5\omega_{n}+b} \quad \forall n
\label{alpha-update}
\end{align}
We see that the solution (\ref{alpha-update}) provides a simple
rule for the update of $\boldsymbol{\alpha}$. Also, notice that
the update rule (\ref{alpha-update}) resembles that of the
conventional sparse Bayesian learning work \cite{Tipping01} except
that $\omega_{n}$ is equal to $\langle x_n^2\rangle$ for the
conventional sparse Bayesian learning method, while for our case,
$\omega_{n}$ is a weighted summation of $\langle x_n^2\rangle$ and
$\langle x_i^2\rangle$ for $i\in N_{(n)}$. Numerical results show
that for a properly chosen value of $a$, this update rule,
although sub-optimal, guarantees an exact recovery and provides
superior recovery performance. Specifically, $a=1.5$ is
empirically proven to be a robust choice which ensures that the
proposed algorithm delivers stable recovery performance.



We now discuss the update of the hyperparameter $\gamma$, the
inverse of the noise variance. Since the GAMP algorithm also
provides an approximate posterior distribution for the noiseless
output $\boldsymbol{z}=\boldsymbol{Ax}$, we can simply treat
$\boldsymbol{z}$ as hidden variables when learning the noise
variance, i.e.
\begin{align}
\gamma^{(t+1)}&=\arg\max_{\gamma}E_{\boldsymbol{z}|\boldsymbol{y},\boldsymbol{\alpha}^{(t)},\gamma^{(t)}}[\log
p(\gamma)p(\boldsymbol{y}|\boldsymbol{z},\gamma)] \nonumber \\
&\triangleq\arg\max_{\gamma}Q(\gamma|\boldsymbol{\alpha}^{(t)},\gamma^{(t)})
\end{align}
Taking the partial derivative of the Q-function with respect to
$\gamma$ gives
\begin{align}
\frac{\partial
Q(\gamma|\boldsymbol{\alpha}^{(t)},\gamma^{(t)})}{\partial\gamma}
=\frac{c-1}{\gamma}-d+\frac{M}{2\gamma}-\frac{1}{2}\sum_{m=1}^{M}\langle(y_m-z_m)^2\rangle
\end{align}
where $\langle\cdot\rangle$ denotes the expectation with respect
to
$p(z_m|\boldsymbol{y},\hat{p}_m(k_0),\tau_m^p(k_0),\boldsymbol{\theta}^{(t)})$,
i.e. the approximate posterior distribution of $z_m$. Recalling
that the approximate posterior of $z_m$ follows a Gaussian
distribution with its mean and variance given by
(\ref{z-post-mean})--(\ref{z-post-var}), we have
\begin{align}
\langle(y_m-z_m)^2\rangle=(y_m-\mu^z_m)^2+\phi_m^z
\end{align}
where $\mu^z_m$ and $\phi_m^z$ are given by
(\ref{z-post-mean})--(\ref{z-post-var}), with
$\{\hat{p}_m,\tau_m^p\}$ replaced by
$\{\hat{p}_m(k_0),\tau_m^p(k_0)\}$, and $\gamma$ replaced by
$\gamma^{(t)}$. Setting the derivative equal to zero, we obtain
the update rule for $\gamma$ as
\begin{align}
\gamma^{(t+1)}=\frac{M+2c-2}{2d+\sum_{m}\langle(y_m-z_m)^2\rangle}
\label{gamma-update}
\end{align}

So far we have completed the development of our GAMP-based
pattern-coupled sparse Bayesian learning algorithm. For clarify,
we now summarize our proposed PCSBL-GAMP algorithm as follows.

\begin{center}
\textbf{PCSBL-GAMP Algorithm}
\end{center}
\vspace{0cm} \noindent
\begin{tabular}{lp{7.7cm}}
\hline 1.& Initialization: given $\boldsymbol{\alpha}^{(0)}$ and $\gamma^{(0)}$.\\
2.& For $t\geq 0$: given $\boldsymbol{\alpha}^{(t)}$ and
$\gamma^{(t)}$, call the GAMP algorithm. Based on the outputs of
the GAMP algorithm, update the hyperparameters
$\boldsymbol{\alpha}^{(t+1)}$ and
$\gamma^{(t+1)}$ according to (\ref{alpha-update}) and (\ref{gamma-update}).\\
3.& Continue the above iteration until the difference between two
consecutive estimates of $\boldsymbol{x}$
is negligible, or a maximum number of iterations is reached.\\
\hline
\end{tabular}

\section{Simulation Results} \label{sec:simulation}
We now carry out experiments to illustrate the performance of our
proposed algorithm, also referred to as PCSBL-GAMP algorithm, and
its comparison with other existing methods. The performance of the
proposed algorithm\footnote{Matlab codes for our algorithm are
available at http://www.junfang-uestc.net/codes/PCSBL-GAMP.rar}
will be examined using both synthetic and real data. The
parameters $a$, $b$, $c$, $d$ for our proposed algorithm are set
equal to $a=1.5$, $b=10^{-6}$, $c=1$, and $d=10^{-6}$ throughout
our experiments. The relevance parameter $\beta$ is set equal to
$\beta=1$ in our experiments. In fact, our empirical results
suggest that its choice is not very critical to the recovery
performance as long as $\beta$ is set into the region $\beta\in
[0.1,1]$.

\subsection{Synthetic Data}
\begin{figure}[!h]
\centering
\includegraphics[width=8cm]{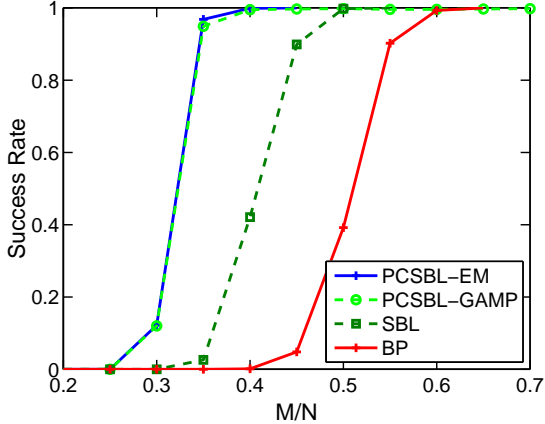}
\caption{Success rates vs. the ratio $M/N$.} \label{fig1}
\end{figure}

\begin{figure}[!h]
\centering
\includegraphics[width=8cm]{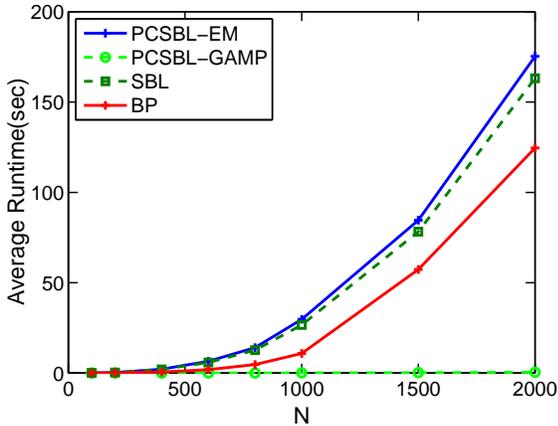}
\caption{Average run times vs. $N$.} \label{fig2}
\end{figure}

We first evaluate the recovery performance of the PCSBL-GAMP
method using the synthetic data. In our simulations, we generate a
one-dimensional block-sparse signal in a way similar to
\cite{FangShen15}. Suppose the $N$-dimensional sparse signal
contains $K$ nonzero coefficients ($K$ is also denoted as the
sparsity level) which are partitioned into $T$ blocks with random
sizes and random locations. The nonzero coefficients of the sparse
signal $\boldsymbol{x}$ and the measurement matrix
$\boldsymbol{A}\in\mathbb{R}^{M\times N}$ are randomly generated
with each entry independently drawn from a normal distribution,
and then the sparse signal $\boldsymbol{x}$ and columns of
$\boldsymbol{A}$ are normalized to unit norm. Fig. \ref{fig1}
depicts the success rate of the proposed PCSBL-GAMP method vs. the
ratio $M/N$, where we set $N=200$, $K=40$ and $T=6$. The success
rate is computed as the ratio of the number of successful trials
to the total number of independent runs. A trial is considered
successful if the normalized squared recovery error
$\|\boldsymbol{x}-\boldsymbol{\hat{x}}\|^2/\|\boldsymbol{x}\|^2$
is no greater than $10^{-6}$, where $\boldsymbol{\hat{x}}$ denotes
the estimate of the sparse signal. The success rates of the
EM-based PC-SBL method \cite{FangShen15} (referred to as the
PCSBL-EM), the conventional SBL \cite{Tipping01}, and the basis
pursuit (BP) method \cite{ChenDonoho98,CandesTao05} are also
included for comparison. From Fig. \ref{fig1}, we see that the
PCSBL-GAMP method achieves almost the same performance as that of
the PCSBL-EM method, and presents a significant performance
advantage over the SBL and BP methods due to exploiting the
underlying block-sparse structures. The average run times of
respective algorithms as a function of the signal dimension $N$ is
plotted in Fig. \ref{fig2}, where we set $M=0.4N$. Results are
averaged over 100 independent runs. We see that the PCSBL-GAMP
requires much less run time than the PCSBL-EM, particularly when
the signal dimension $N$ is large. Also, it can be observed that
the average run time of the PCSBL-EM grows rapidly with an
increasing $N$, whereas the average run time of the PCSBL-GAMP
increases very slowly. This observation coincides with our
computational complexity analysis.


\begin{figure*}[!t]
 \centering
\begin{tabular}{ccccccc}
\hspace*{-2ex}
\includegraphics[width=3.3cm]{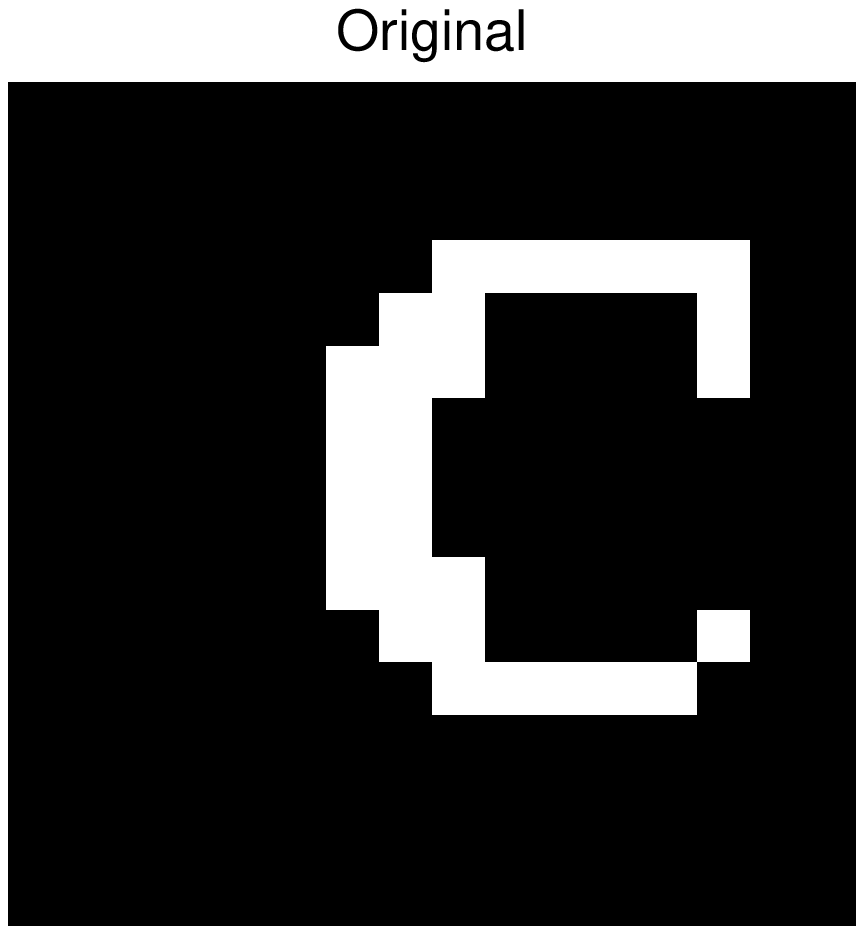} &
\hspace*{-2ex}
\includegraphics[width=3.3cm]{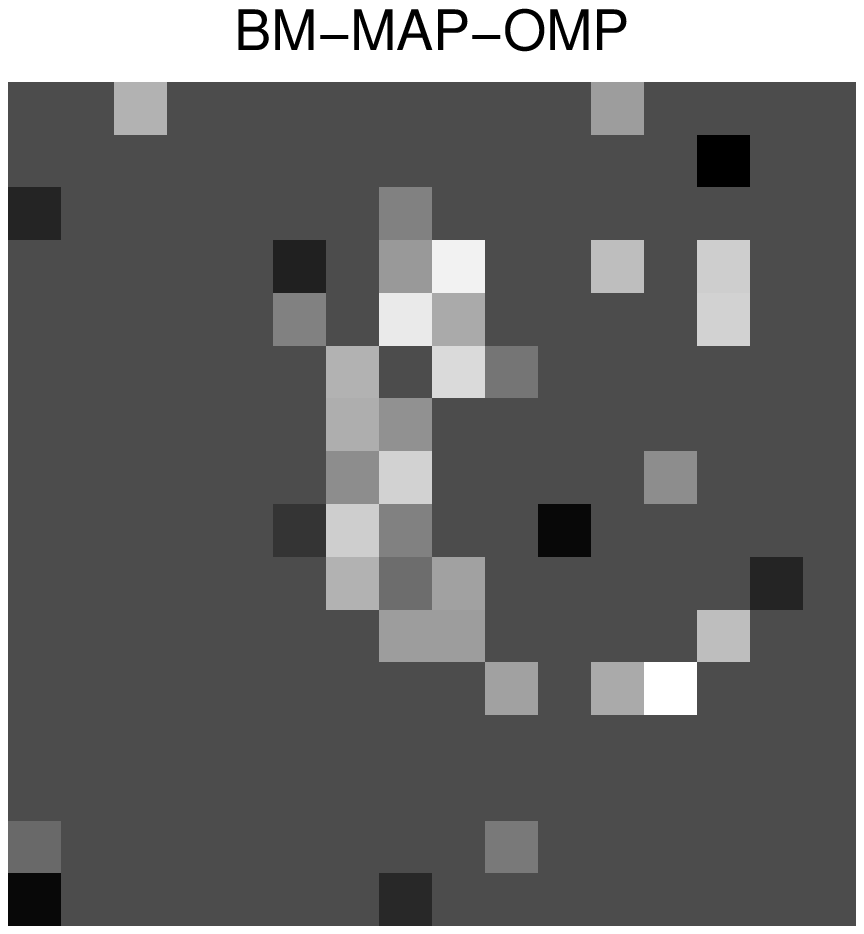} &
\hspace*{-2ex}
\includegraphics[width=3.3cm]{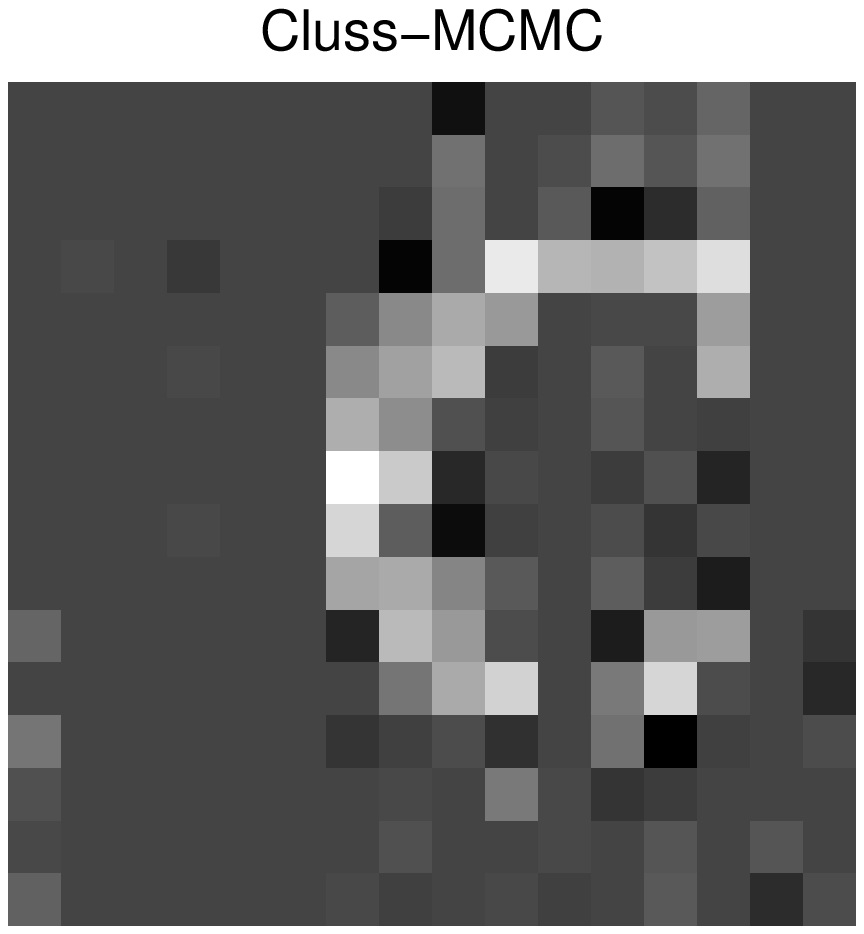} &
\hspace*{-2ex}
\includegraphics[width=3.3cm]{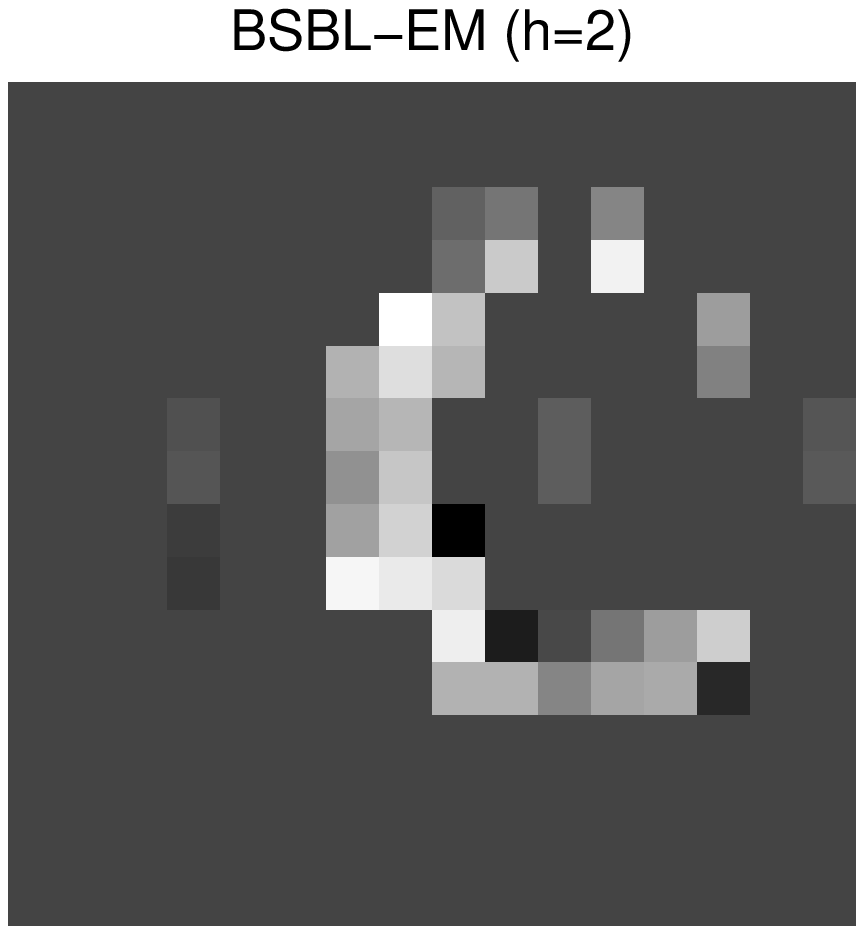} &
\hspace*{-2ex}
\includegraphics[width=3.3cm]{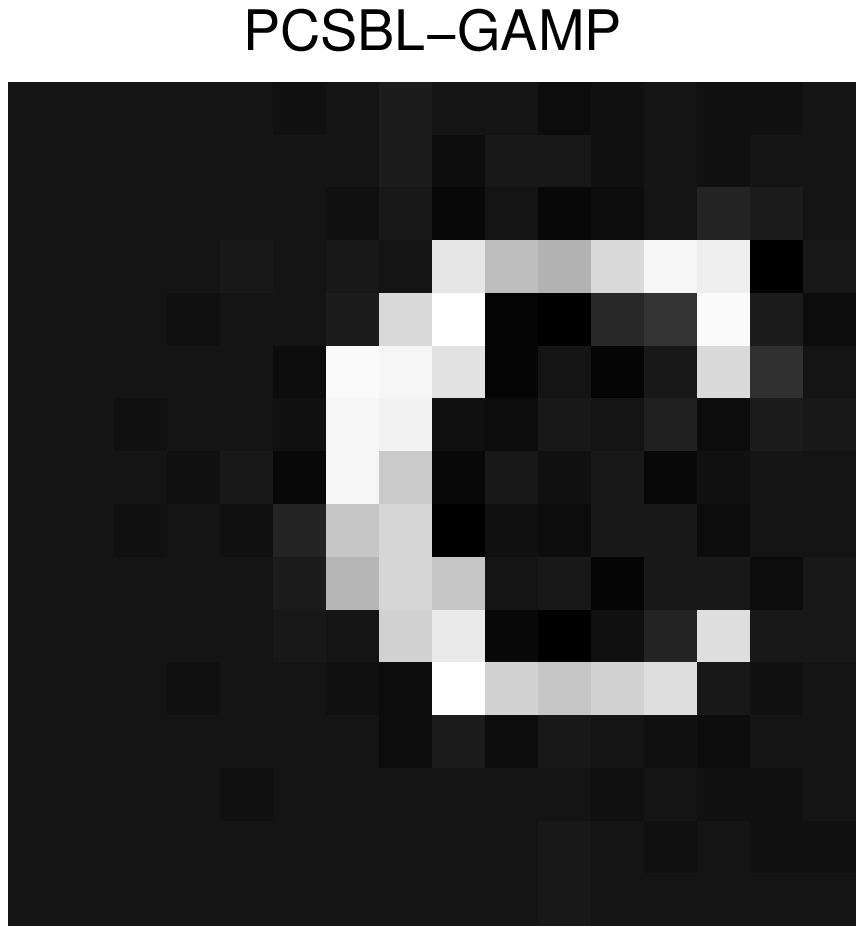} \\
\hspace*{-2ex}
\includegraphics[width=3.3cm]{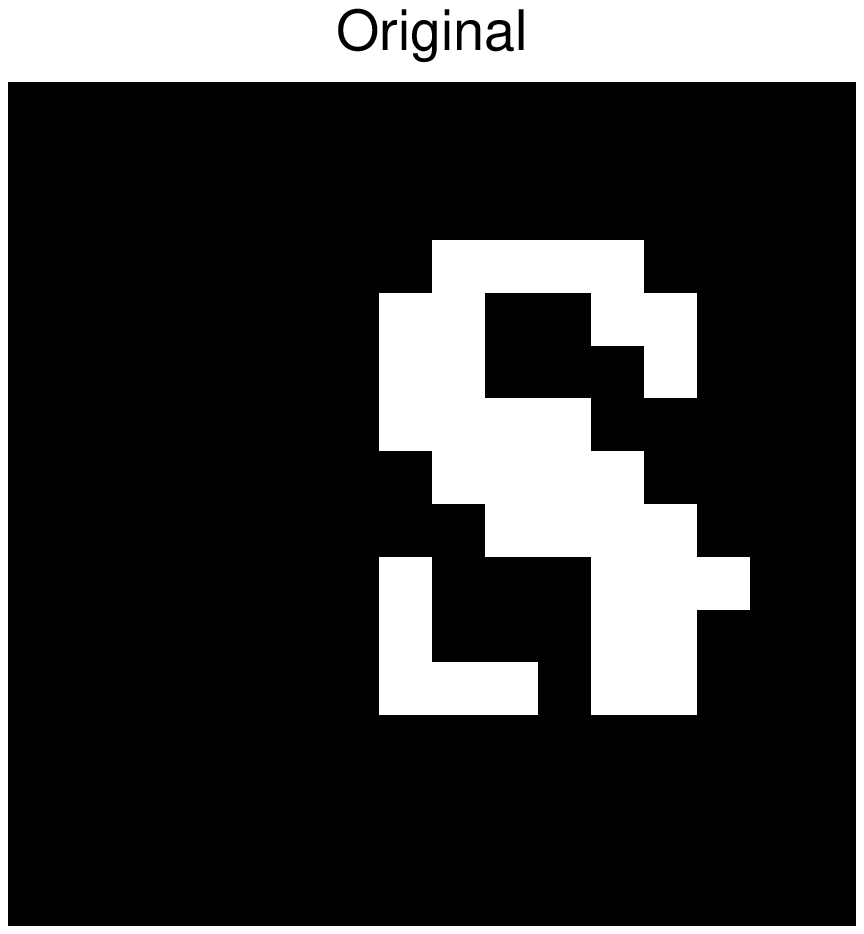} &
\hspace*{-2ex}
\includegraphics[width=3.3cm]{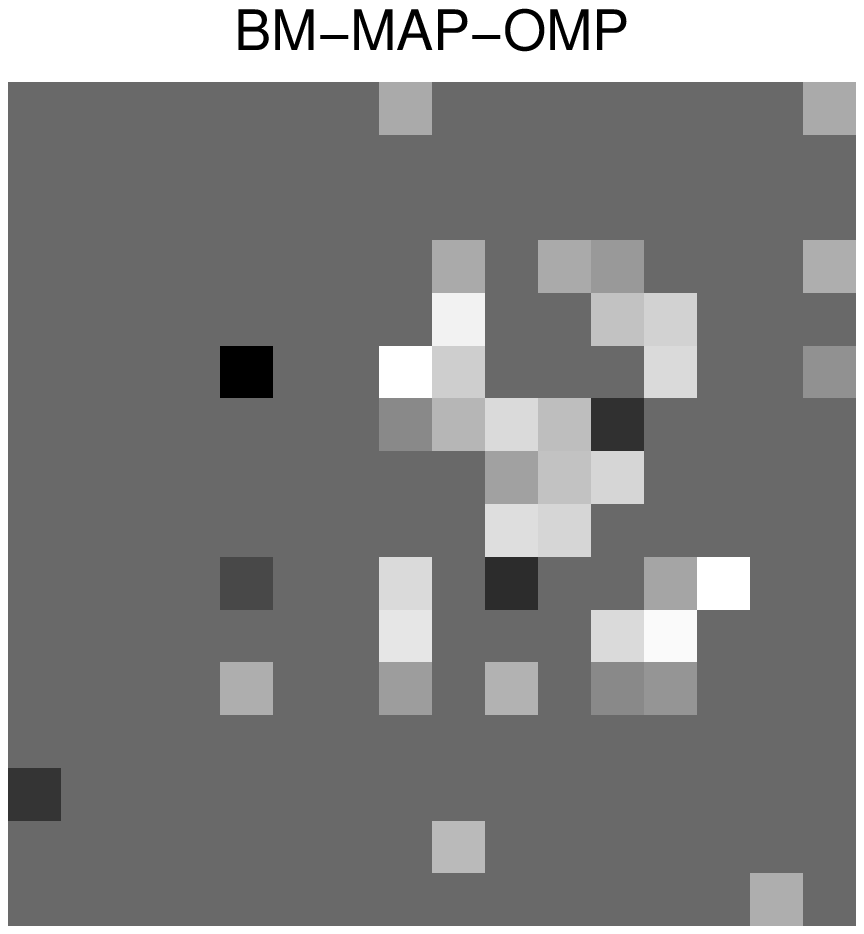} &
\hspace*{-2ex}
\includegraphics[width=3.3cm]{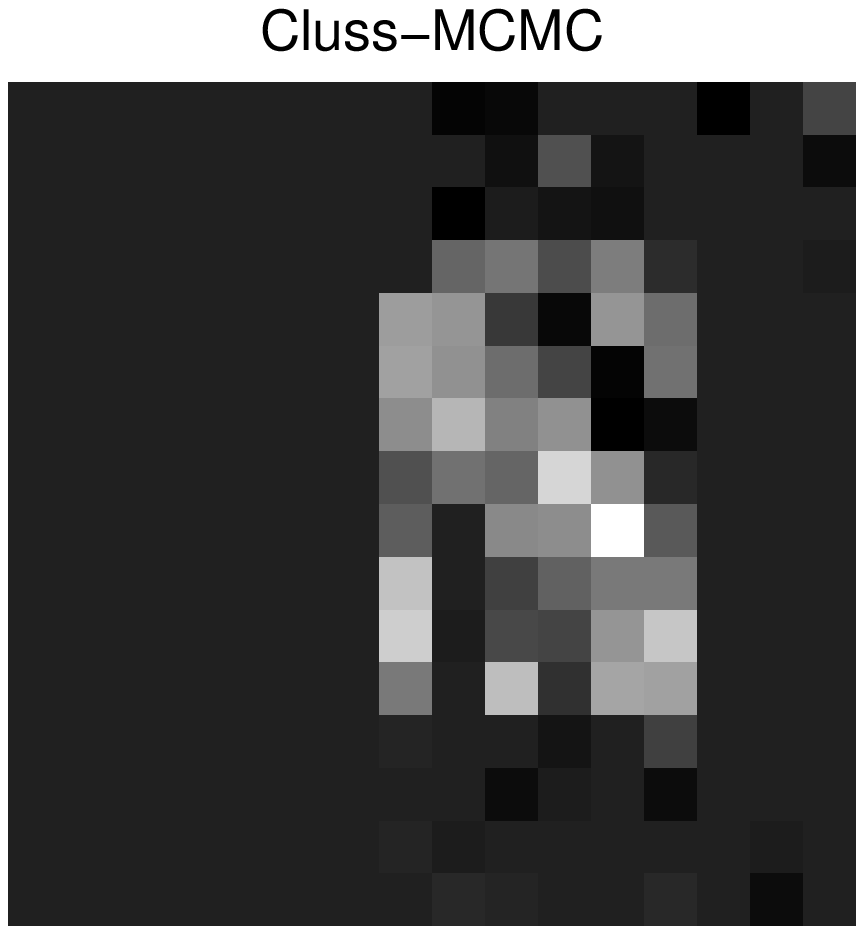}&
\hspace*{-2ex}
\includegraphics[width=3.3cm]{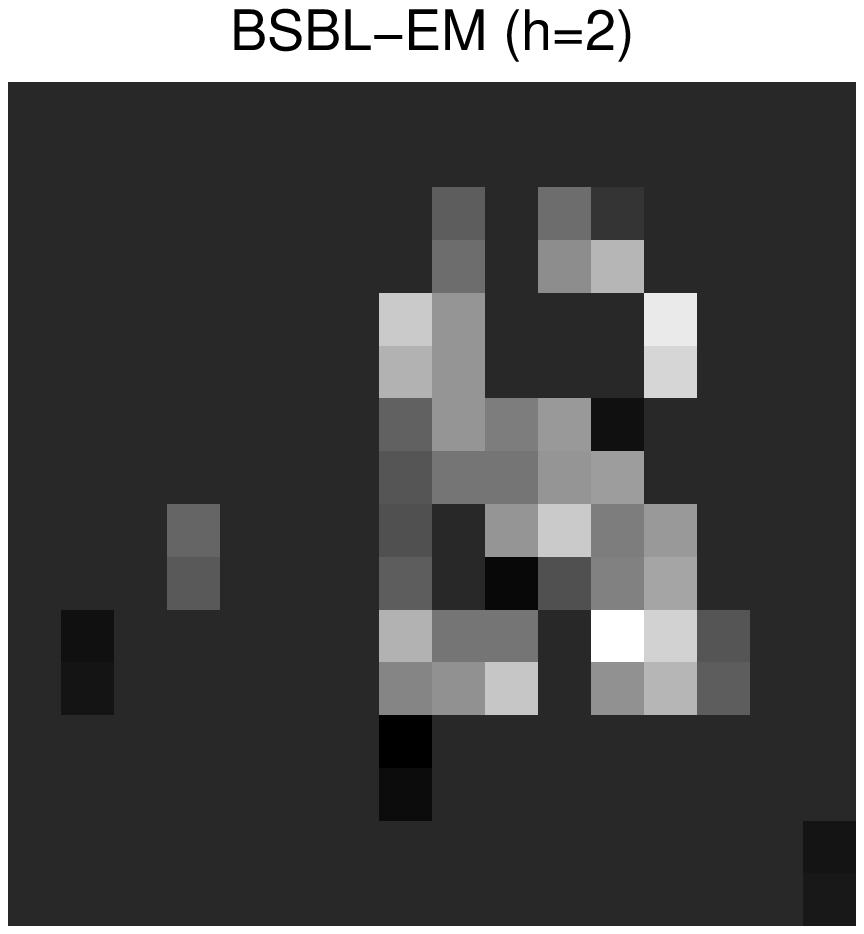} &
\hspace*{-2ex}
\includegraphics[width=3.3cm]{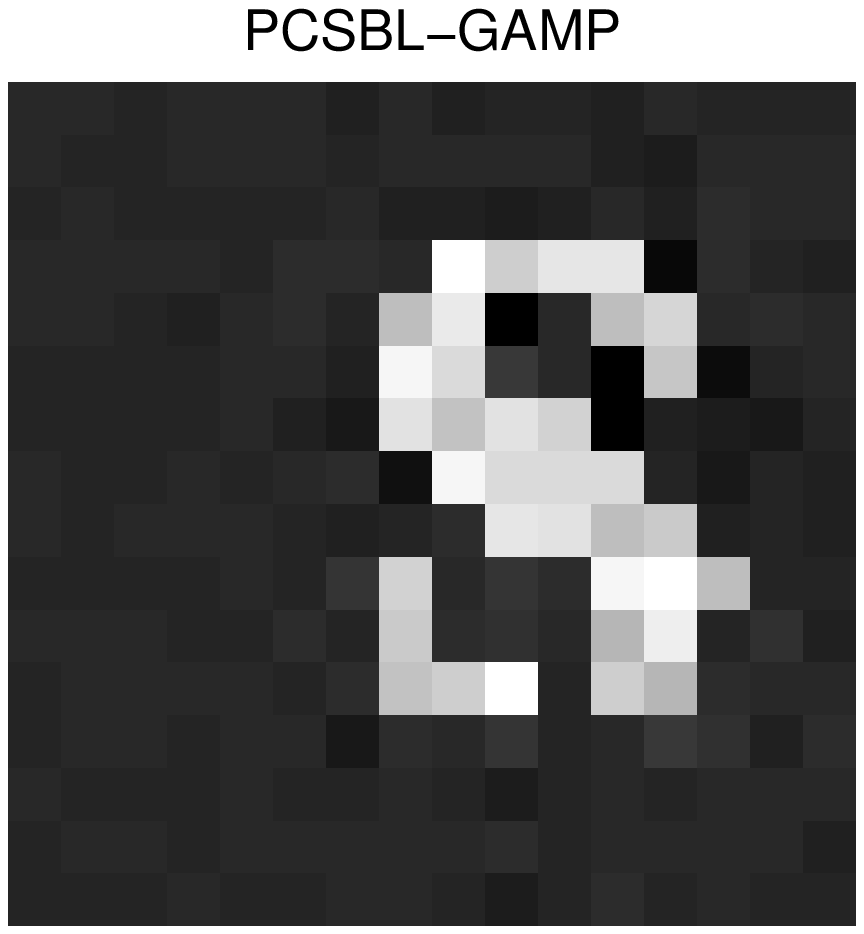}
\end{tabular}
  \caption{Original patches of letters and patches reconstructed by respective algorithms.}
   \label{fig3}
\end{figure*}

We also carry out experiments using patches of letters ``C'' and
``S'' ($16\times 16$ pixels) with black background, where most of
the pixels on the patches are zeros and the nonzero coefficients
exhibit irregular block patterns. White Gaussian noise is added to
the patches. We compare our method with other block-sparse signal
recovery algorithms, namely, the cluster-structured MCMC algorithm
(Cluss-MCMC) \cite{YuSun12}, the Boltzman machine-based greedy
pursuit algorithm (BM-MAP-OMP) \cite{PelegEldar12}, and the block
sparse Bayesian learning method (BSBL) method
\cite{ZhangRao11,ZhangRao13}. Note that although the BM-MAP-OMP
and the BSBL algorithms are developed for one-dimensional sparse
signal recovery, we extend their methods to the two-dimensional
scenario. In our simulations, model parameters used by the
competing algorithms are adjusted to achieve the best performance.
For the BSBL algorithms, the block size parameter $h$ is set equal
to 2. Fig. \ref{fig3} depicts the original patches and the patches
reconstructed by respective algorithms, where we set $M=80$, and
the signal-to-noise ratio (SNR) is set to 20dB with
$\text{SNR}\triangleq 10\log(\|\boldsymbol{Ax}\|_2^2/M\sigma^2)$.
It can be seen that our proposed PCSBL-GAMP method provides the
best visual quality with recognizable letters, whereas the letters
reconstructed by other algorithms have considerably lower quality,
particularly for the Cluss-MCMC and the BM-MAP-OMP methods. This
result also implies that our proposed method is flexible to
accommodate any irregular cluster patterns.



\begin{figure*}[!t]
 \centering
\begin{tabular}{ccccc}
\hspace*{-10ex}
\includegraphics[width=4.5cm]{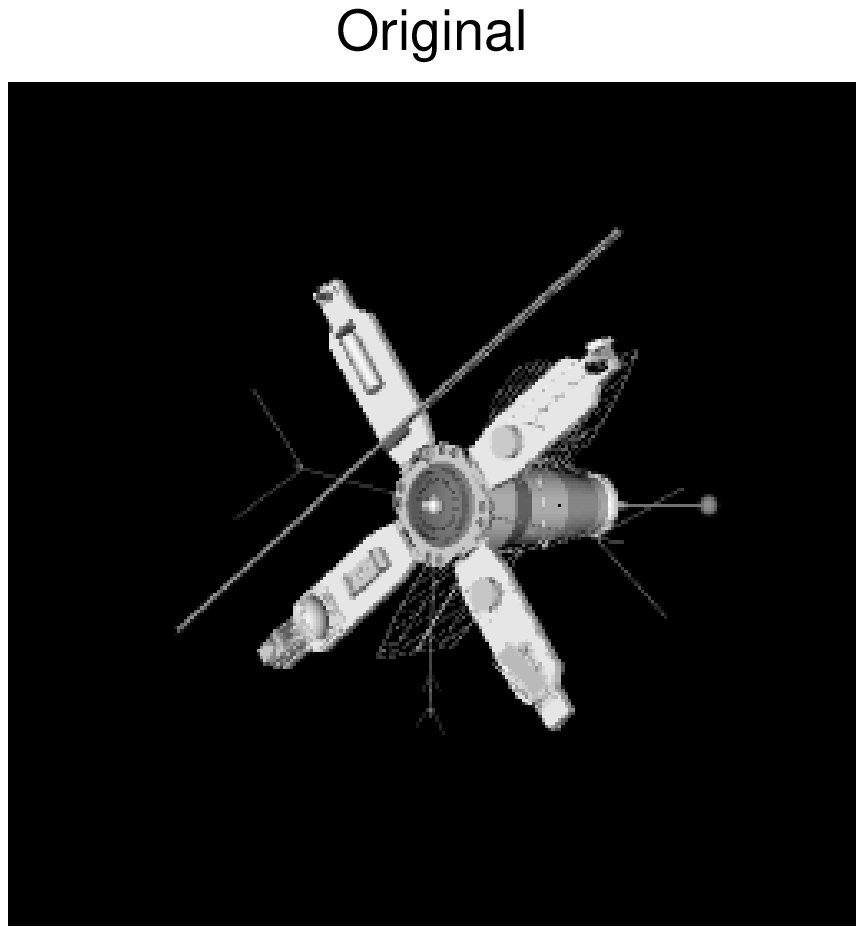} &
\hspace*{-6ex}
\includegraphics[width=4.5cm]{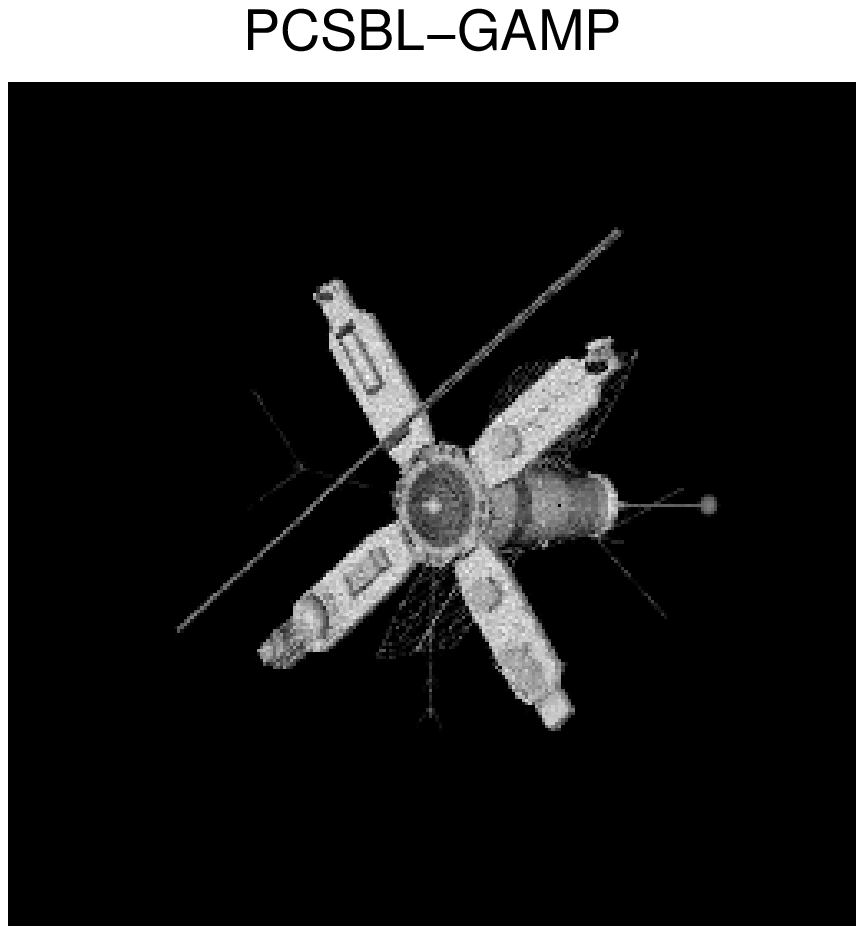} &
\hspace*{-6ex}
\includegraphics[width=4.5cm]{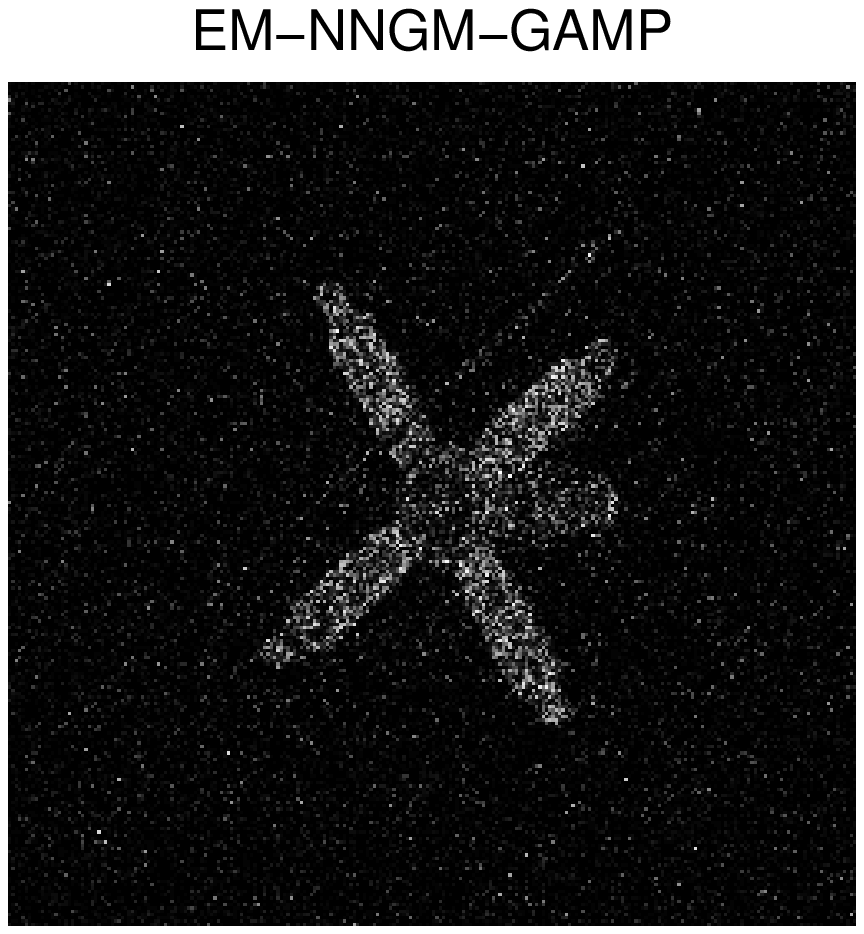}
& \hspace*{-6ex}
\includegraphics[width=4.5cm]{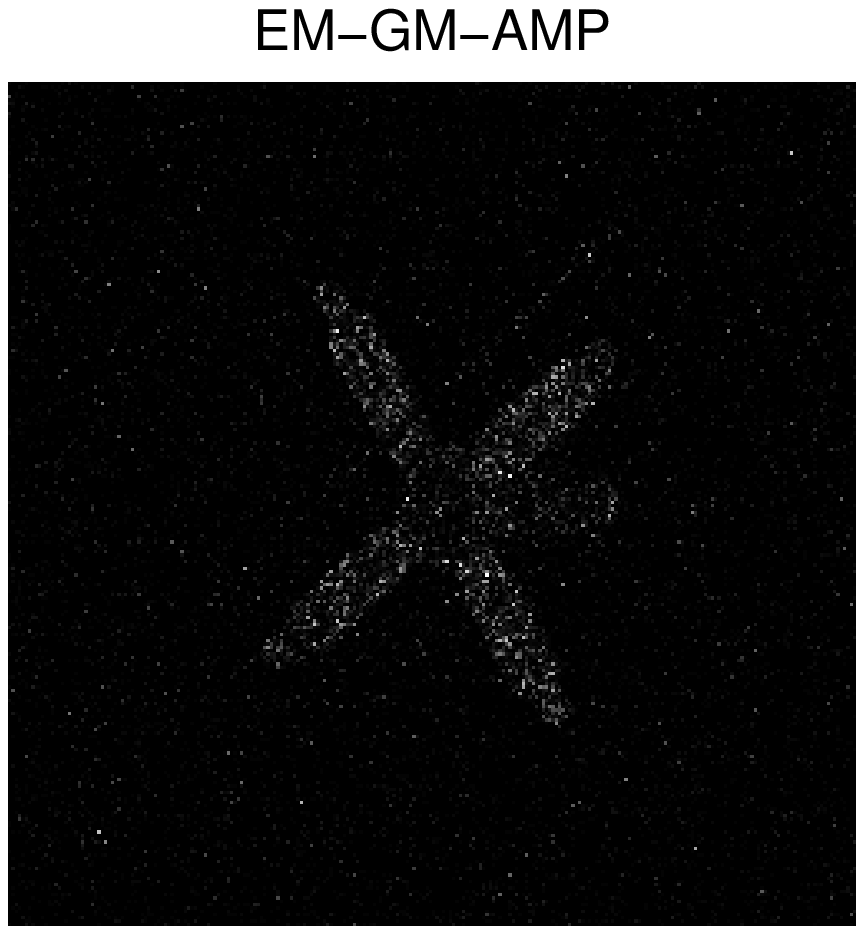}
& \hspace*{-6ex}
\includegraphics[width=4.5cm]{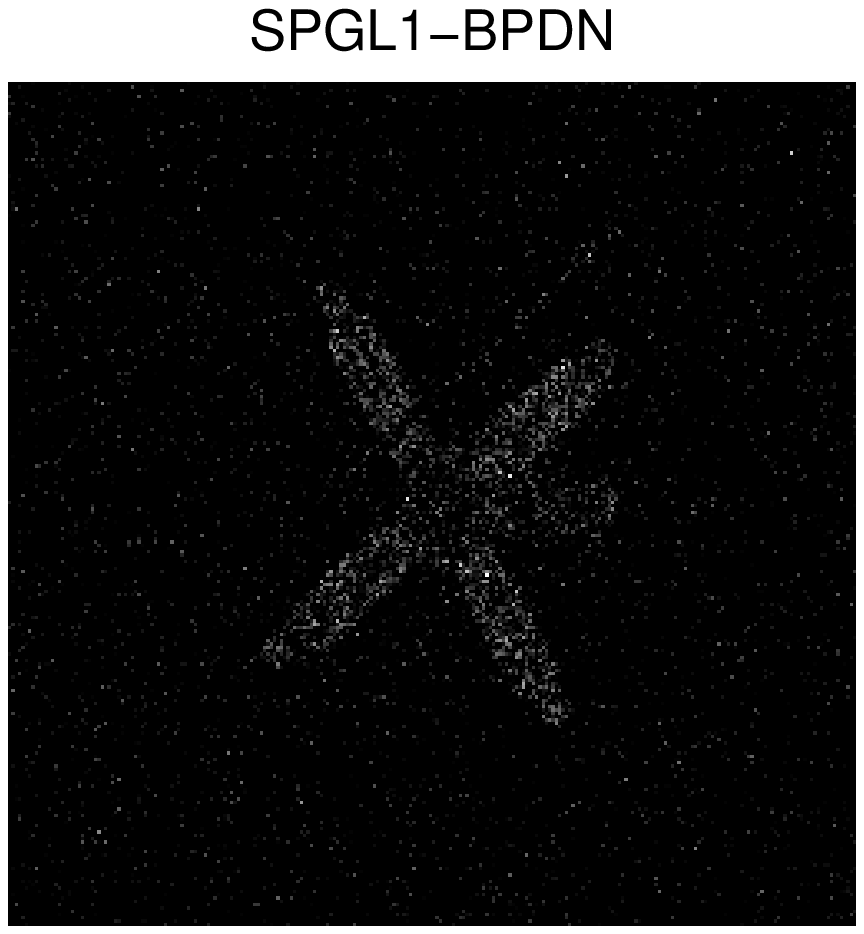}
\end{tabular}
  \caption{Original satellite image and images reconstructed by respective algorithms.}
   \label{fig4}
\end{figure*}

\subsection{Satellite Image Recovery}
\begin{figure}[!h]
\centering
\includegraphics[width=8cm]{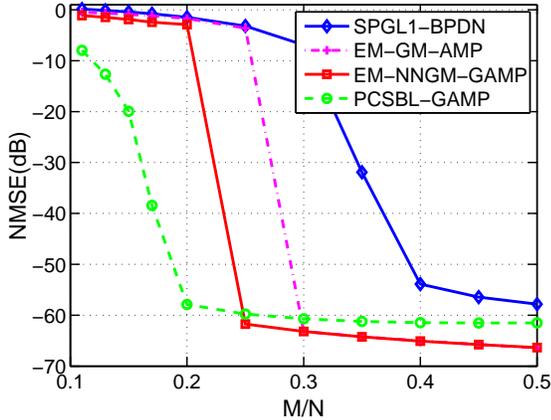}
\caption{Satellite image: NMSEs vs. the ratio $M/N$.} \label{fig5}
\end{figure}

In this subsection, we carry out experiments on a non-negative
$256\times 256$ satellite image\footnote{Image data is available
at http://sourceforge.net/projects/gampmatlab.}. The image is
sparse in the spatial domain, with only $6678$ (approximately
$10.2\%$ of total pixels) nonzero pixels. In our experiments,
compressive measurements are corrupted by additive i.i.d. Gaussian
noise, i.e. $\boldsymbol{y}=\boldsymbol{Ax}+\boldsymbol{w}$, where
the image is represented as a one-dimensional vector
$\boldsymbol{x}$. The sensing matrix $\boldsymbol{A}$ is chosen to
be the same as that used in \cite{VilaSchniter14}, i.e.
$\boldsymbol{A}=\boldsymbol{\Phi\Psi S}$, where
$\boldsymbol{\Phi}\in\{0,1\}^{M\times N}$ and its rows are
randomly selected from the $N\times N$ identity matrix,
$\boldsymbol{\Psi}\in\{-1,1\}^{N\times N}$ is a Hadamard transform
matrix, $\boldsymbol{S}\in\mathbb{R}^{N\times N}$ is a diagonal
matrix with its entries randomly chosen from $\{-1,1\}$. Sensing
using such a measurement matrix can be executed using a fast
binary algorithm, which makes the hardware implementation simple.
Note that the BM-MAP-OMP, the BSBL and the Cluss-MCMC methods were
not included in this experiment due to their prohibitive
computational complexity when the signal dimension is large.
Instead, we compare our method with some other computationally
efficient GAMP-based methods, namely, the EM-NNGM-GAMP method
\cite{VilaSchniter14} and the EM-GM-AMP method
\cite{VilaSchniter13}. These two methods have demonstrated
state-of-the-art recovery performance in a series of experiments.
The spectral projected gradient (SPG) method (referred to as
SPGL1) which was developed in \cite{BergFriedlander08} to
efficiently solve the basis pursuit or basis pursuit denoising
optimizations is also included for comparison. Among these
methods, only the EM-NNGM-GAMP algorithm exploits the
non-negativity of the satellite image. Therefore the signals
recovered by these algorithms, except the EM-NNGM-GAMP, may
contain negative coefficients. These negative coefficients are
manually set to zero in our simulations. Fig. \ref{fig4} shows the
original satellite image and the images reconstructed by
respective algorithms, where we set $M=0.15N$ and
$\text{SNR}=60\text{dB}$. It can be seen that the PCSBL-GAMP
offers a significantly better image quality as compared with other
methods. Fig. \ref{fig5} plots the normalized mean square errors
(NMSEs) of respective algorithms vs. the ratio $M/N$. Results are
averaged over 100 independent trails, with the sensing matrix
$\boldsymbol{A}$ randomly generated for each trial. The recovered
negative coefficients are kept unaltered in calculating the NMSEs.
We see that our proposed PCSBL-GAMP method outperforms other
algorithms by a big margin for a small ratio $M/N$ (e.g. $M/N\leq
0.25$), where data acquisition is practically appealing due to
high compression rates.

\subsection{Background Subtraction}
\begin{figure*}[!t]
 \centering
\begin{tabular}{ccc}
\hspace*{-10ex}
\includegraphics[width=4.5cm]{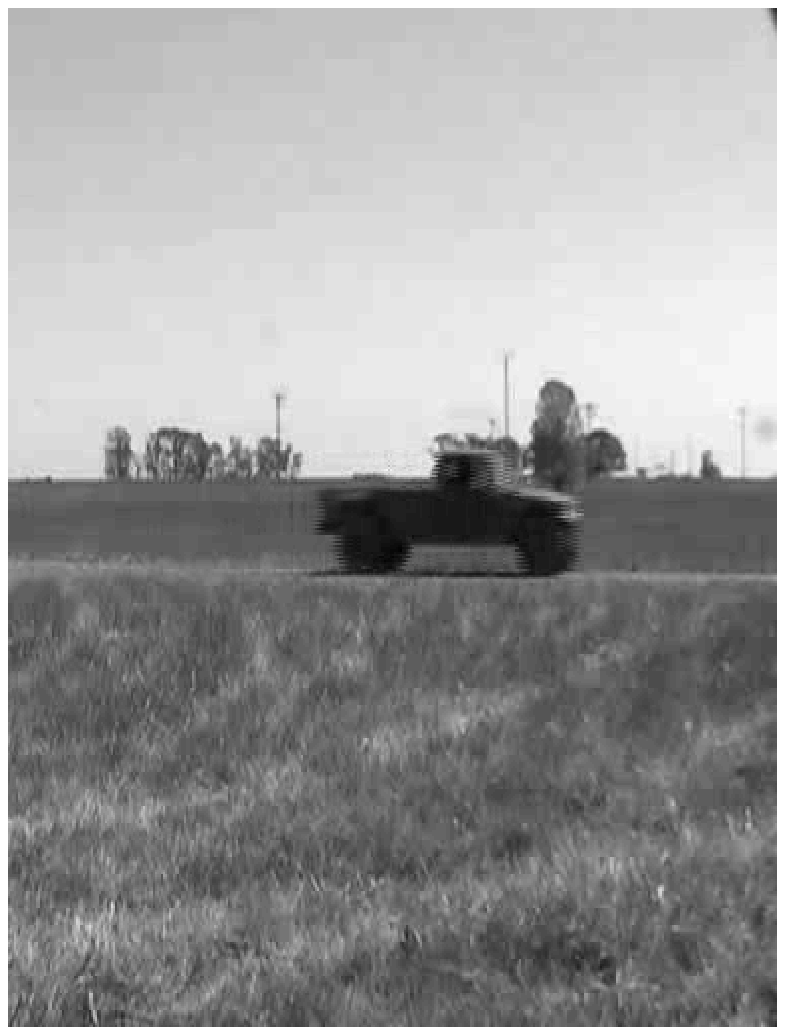} &
\hspace*{-0ex}
\includegraphics[width=4.5cm]{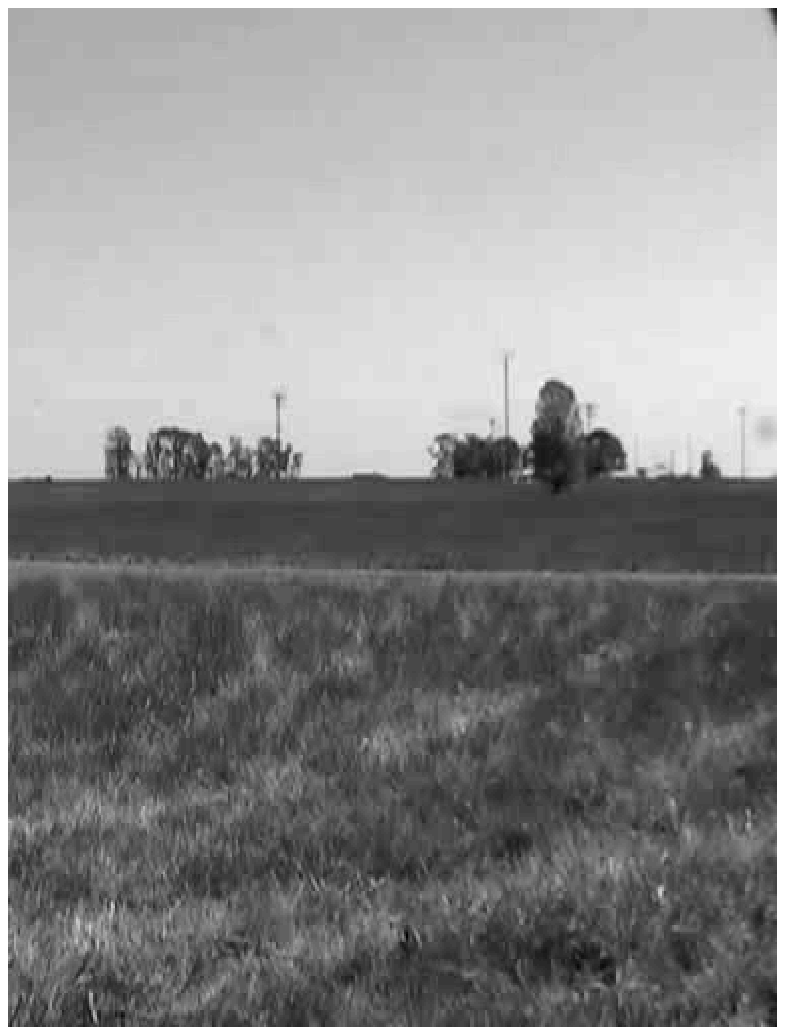} &
\hspace*{-0ex}
\includegraphics[width=4.5cm]{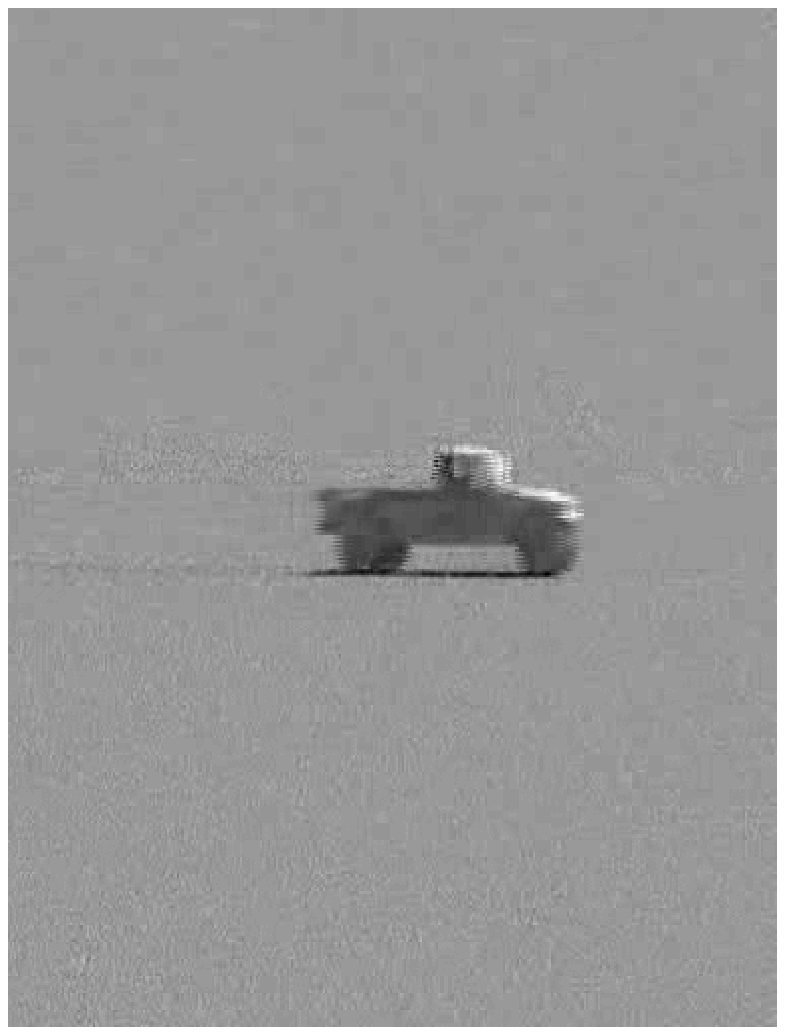}
\end{tabular}
  \caption{From left to
    right: the test image $\boldsymbol{x}_t$ (40th frame of the Convoy2 data set), the background image
    $\boldsymbol{x}_b$,
    the foreground image $\boldsymbol{x}_f$.}
   \label{fig6}
\end{figure*}

\begin{figure*}[!t]
 \centering
\begin{tabular}{cccc}
\hspace*{-5ex}
\includegraphics[width=4.5cm]{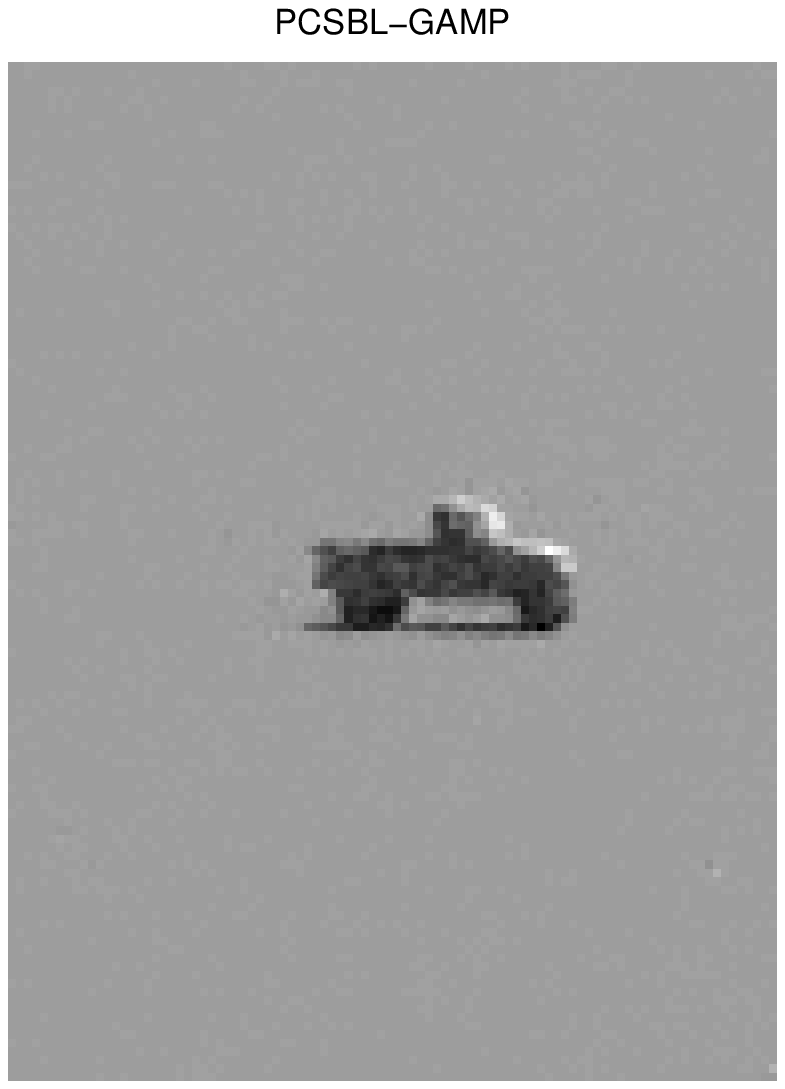} &
\hspace*{-0ex}
\includegraphics[width=4.5cm]{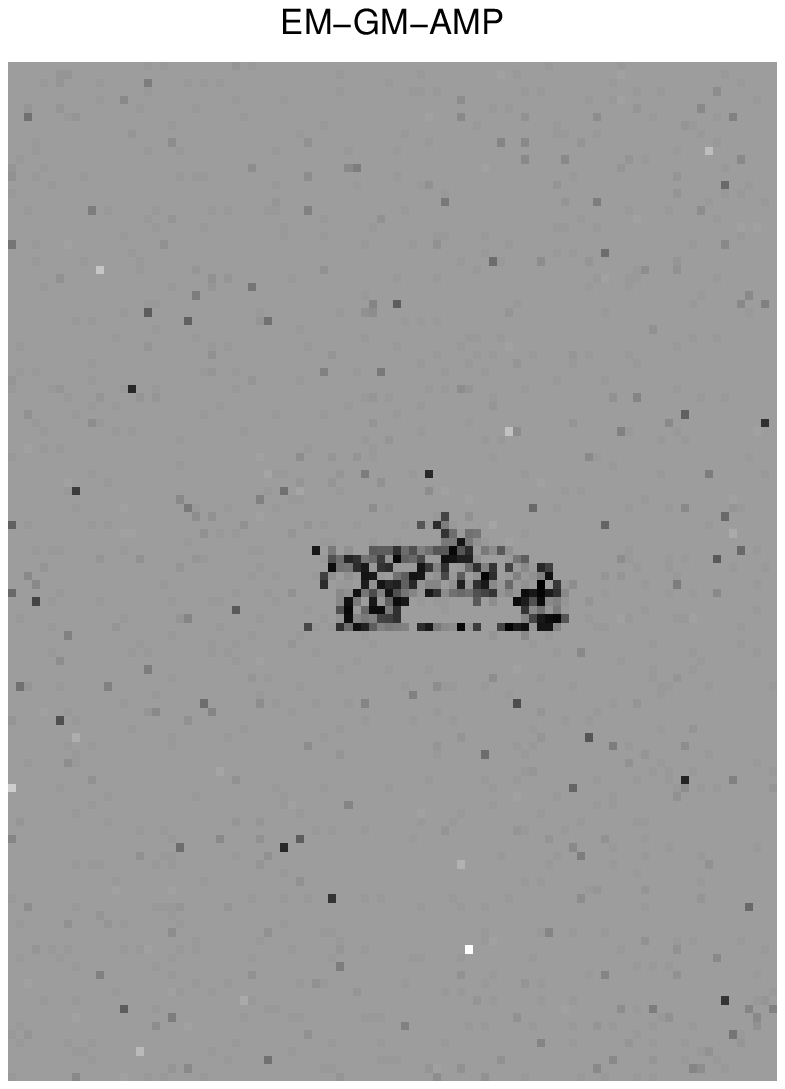} &
\hspace*{-0ex}
\includegraphics[width=4.5cm]{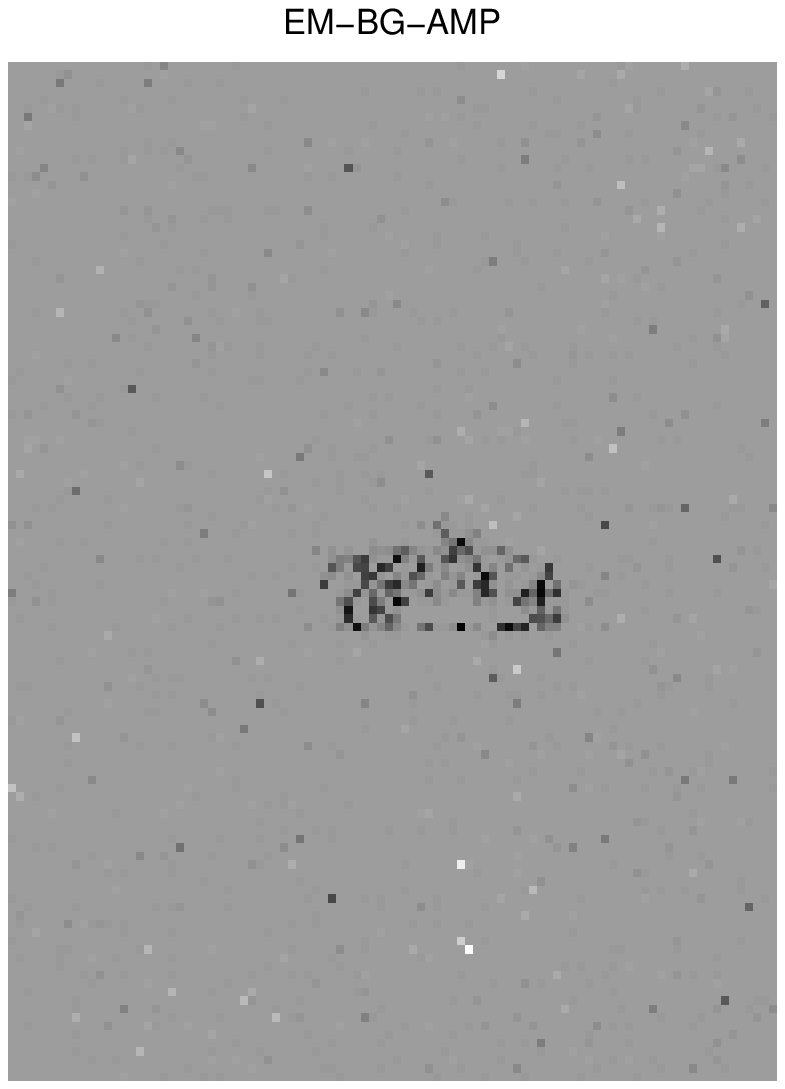}
& \hspace*{-0ex}
\includegraphics[width=4.5cm]{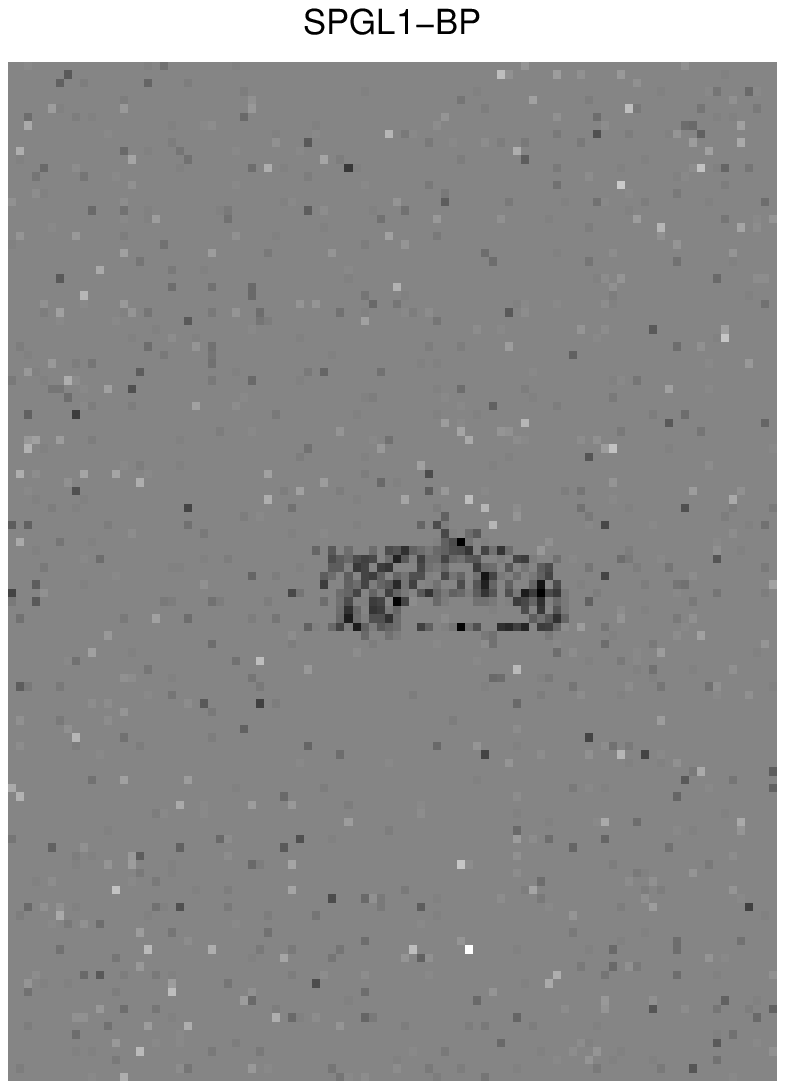}
\end{tabular}
  \caption{Foreground images reconstructed by respective algorithms.}
   \label{fig7}
\end{figure*}

\begin{figure}[!h]
\centering
\includegraphics[width=8cm]{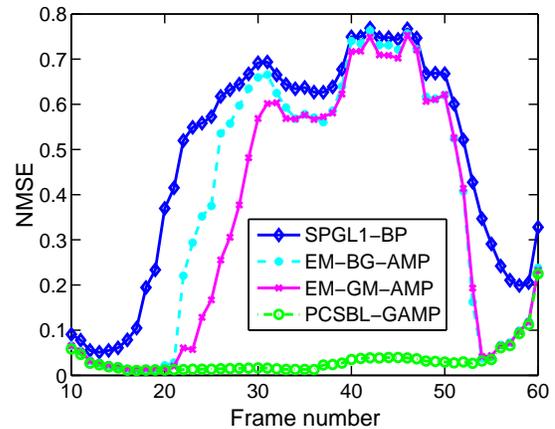}
\caption{NMSEs vs. the frame number.} \label{fig8}
\end{figure}

Background subtraction, also known as foreground detection, is a
technique used to automatically detect and track moving objects in
videos from static cameras. Usually, the foreground innovations
are sparse in the spatial image domain. By exploiting this
sparsity, the sparse foreground innovations within a scene can be
reconstructed using compressed measurements, which relieves the
communications burden placed on data transmission \cite{Cevher11}.
Specifically, the idea is to reconstruct the foreground image from
the difference between the compressed measurements of the
background image and the compressed measurements of the test image
\cite{Cevher11}
\begin{align}
\boldsymbol{y}_f=&\boldsymbol{y}_t-\boldsymbol{y}_b=\boldsymbol{A}(\boldsymbol{x}_t-\boldsymbol{x}_b)
\nonumber\\
=&\boldsymbol{A}\boldsymbol{x}_f \nonumber
\end{align}
where $\boldsymbol{x}_t$ and $\boldsymbol{x}_b$ represent the test
and the background images, respectively; $\boldsymbol{y}_t$ and
$\boldsymbol{y}_b$ denote the compressed measurements of the test
and background images, respectively; and $\boldsymbol{x}_f$ is the
foreground image to be recovered. In our experiments, we use the
Convoy2 data set that was used in \cite{WarnellReddy12}. The
Convoy2 data set was collected on the Spesutie island, consisting
of a video sequence with $260$ frames and one background frame
recorded by a single static camera. The video sequence has a
dynamic sparse foreground as vehicles enter and exit the filed of
view over time. We first choose the $40$th frame of the Convoy2
data set as a test image, which is shown in Fig. \ref{fig6}. The
background image and the foreground image are also included in
Fig. \ref{fig6}. The foreground image is regarded as the
groundtruth image. This foreground image, however, does not have a
pure background since
$\boldsymbol{x}_f=\boldsymbol{x}_t-\boldsymbol{x}_b$ is not an
exactly sparse signal and contains many small nonzero components.
In our experiments, the original images of $480\times 381$ pixels
are resized to $120\times 96$ pixels. For the resized foreground
image, we have a total number of $923$ coefficients whose
magnitudes are greater than $10^{-2}$, thus the percentage of
nonzero coefficients is $923/(120\times 96)=8.01\%$. Again, the
BM-MAP-OMP, BSBL, and the Cluss-MCMC methods are not included due
to their prohibitive computational complexity. Here we compare our
method with the EM-GM-AMP method \cite{VilaSchniter13} and the
EM-BG-AMP method \cite{VilaSchniter11}. Fig. \ref{fig7} depicts
images reconstructed by respective algorithms, where we use only
$M=0.1N$ measurements. The measurement matrix $\boldsymbol{A}$ is
randomly generated with each entry independently drawn from a
normal distribution. We see that our proposed PCSBL-GAMP method
provides the finest image quality with a clear appearance of the
vehicle, whereas the object silhouettes recovered by other methods
are hardly recognizable. In our next experiments, frames from the
10th to 60th are used as test images. For each test image, we use
$M=0.1N$ measurements to recover the difference (foreground)
image. Fig. \ref{fig8} shows the NMSEs of respective algorithms
vs. the frame number, where the NMSEs are obtained by averaging
over 100 independent runs, with the measurement matrix
$\boldsymbol{A}$ randomly generated for each run. From Fig.
\ref{fig8}, we observe that our proposed method presents a
significant performance advantage over other methods.

\section{Conclusions} \label{sec:conclusions}
We developed a pattern-coupled sparse Bayesian learning method for
recovery of two-dimensional block-sparse signals whose cluster
patterns are unknown \emph{a priori}. A two-dimensional
pattern-coupled hierarchical Gaussian prior model is introduced to
characterize and exploit the pattern dependencies among
neighboring coefficients. The proposed pattern-coupled
hierarchical model is effective and flexible to capture any
underlying block-sparse structures, without requiring the prior
knowledge of the block partition. An expectation-maximization (EM)
strategy is employed to infer the maximum a posterior (MAP)
estimate of the hyperparameters, along with the posterior
distribution of the sparse signal. Additionally, the generalized
approximate message passing (GAMP) algorithm is embedded in the EM
framework to efficiently compute an approximation of the posterior
distribution of hidden variables, which results in a significant
reduction in computational complexity. Numerical results show that
our proposed algorithm presents a substantial performance
advantage over other existing state-of-the-art methods in image
recovery.


\begin{thebibliography}{10}
\providecommand{\url}[1]{#1} \csname url@rmstyle\endcsname
\providecommand{\newblock}{\relax}
\providecommand{\bibinfo}[2]{#2}
\providecommand\BIBentrySTDinterwordspacing{\spaceskip=0pt\relax}
\providecommand\BIBentryALTinterwordstretchfactor{4}
\providecommand\BIBentryALTinterwordspacing{\spaceskip=\fontdimen2\font
plus \BIBentryALTinterwordstretchfactor\fontdimen3\font minus
  \fontdimen4\font\relax}
\providecommand\BIBforeignlanguage[2]{{%
\expandafter\ifx\csname l@#1\endcsname\relax
\typeout{** WARNING: IEEEtran.bst: No hyphenation pattern has been}%
\typeout{** loaded for the language `#1'. Using the pattern for}%
\typeout{** the default language instead.}%
\else \language=\csname l@#1\endcsname \fi #2}}

\bibitem{TroppGilbert07}
J.~A. Tropp and A.~C. Gilbert, ``Signal recovery from random
measurements via
  orthogonal matching pursuit,'' \emph{IEEE Trans. Information Theory},
  vol.~53, no.~12, pp. 4655--4666, Dec. 2007.

\bibitem{ChenDonoho98}
S.~S. Chen, D.~L. Donoho, and M.~A. Saunders, ``Atomic
decomposition by basis
  pursuit,'' \emph{SIAM J. Sci. Comput.}, vol.~20, no.~1, pp. 33--61, 1998.

\bibitem{WipfNagaranjan10}
D.~Wipf and S.~Nagarajan, ``Iterative reweighted $\ell_1$ and
$\ell_2$ methods
  for finding sparse solutions,'' \emph{IEEE Journals of Selected Topics in
  Signal Processing}, vol.~4, no.~2, pp. 317--329, Apr. 2010.

\bibitem{Tipping01}
M.~Tipping, ``Sparse {B}ayesian learning and the relevance vector
machine,''
  \emph{Journal of Machine Learning Research}, vol.~1, pp. 211--244, 2001.

\bibitem{JiXue08}
S.~Ji, Y.~Xue, and L.~Carin, ``Bayesian compressive sensing,''
\emph{IEEE
  Trans. Signal Processing}, vol.~56, no.~6, pp. 2346--2356, June 2008.

\bibitem{YangXie13}
Z.~Yang, L.~Xie, and C.~Zhang, ``Off-grid direction of arrival
estimation using
  sparse {B}ayesian inference,'' \emph{IEEE Trans. Signal Processing}, vol.~61,
  no.~1, pp. 38--42, Jan. 2013.

\bibitem{WangZhao14}
L.~Wang, L.~Zhao, G.~Bi, C.~Wan, and L.~Yang, ``Enhanced {ISAR}
imaging by
  exploiting the continuity of the target scene,'' \emph{IEEE Trans. Geoscience
  and Remote Sensing}, no.~9, pp. 5736--5750, Sept. 2014.

\bibitem{Cevher11}
V.~Cevher, A.~Sankaranarayanan, M.~F. Duarte, D.~Reddy, R.~G.
Baraniuk, and
  R.~Chellappa, ``Compressive sensing for background subtraction,'' in
  \emph{European Conf. Comp. Vision (ECCV)}, Marseille, France, October 12-18
  2008.

\bibitem{VaswaniLu10}
N.~Vaswani and W.~Lu, ``Modified-{CS}: modifying compressive
sensing for
  problems with partially known support,'' \emph{IEEE Trans. Signal
  Processing}, no.~9, pp. 4595--4607, September 2010.

\bibitem{EldarMishali09}
Y.~C. Eldar and M.~Mishali, ``Robust recovery of signals from a
structured
  union of subspaces,'' \emph{IEEE Trans. Information Theory}, vol.~55, no.~11,
  pp. 5302--5316, Nov. 2009.

\bibitem{BaraniukCevher10}
R.~G. Baraniuk, V.~Cevher, M.~F. Duarte, and C.~Hegde,
``Model-based
  compressive sensing,'' \emph{IEEE Trans. Information Theoy}, vol.~56, no.~4,
  pp. 1982--2001, Apr. 2010.

\bibitem{EldarKuppinger10}
Y.~C. Eldar, P.~Kuppinger, and H.~B\"{o}lcskei, ``Block-sparse
signals:
  uncertainty relations and efficient recovery,'' \emph{IEEE Trans. Information
  Theory}, vol.~58, no.~6, pp. 3042--3054, June 2010.

\bibitem{YuanLin06}
M.~Yuan and Y.~Lin, ``Model selection and estimation in regression
with grouped
  variables,'' \emph{J. R. Statist. Soc. B}, vol.~68, pp. 49--67, 2006.

\bibitem{WipfRao07}
D.~P. Wipf and B.~D. Rao, ``An empirical {B}ayesian strategy for
solving the
  simultaneous sparse approximation problem,'' \emph{IEEE Trans. Signal
  Processing}, vol.~55, no.~7, pp. 3704--3716, July 2007.

\bibitem{ZhangRao11}
Z.~Zhang and B.~D. Rao, ``Sparse signal recovery with temporally
correlated
  source vectors using sparse {B}ayesian learning,'' \emph{IEEE Journal of
  Selected Topics in Signal Processing}, vol.~5, no.~5, pp. 912--926, Sept.
  2011.

\bibitem{HeCarin09}
L.~He and L.~Carin, ``Exploiting structure in wavelet-based
{B}ayesian
  compressive sensing,'' \emph{IEEE Trans. Signal Processing}, vol.~57, no.~9,
  pp. 3488--3497, Sept. 2009.

\bibitem{YuSun12}
L.~Yu, H.~Sun, J.~P. Barbot, and G.~Zheng, ``{B}ayesian
compressive sensing for
  cluster structured sparse signals,'' \emph{Signal Processing}, vol.~92, pp.
  259--269, 2012.

\bibitem{PelegEldar12}
T.~Peleg, Y.~C. Eldar, and M.~Elad, ``Exploiting statistical
dependencies in
  sparse representations for signal recovery,'' \emph{IEEE Trans. Signal
  Processing}, vol.~60, no.~5, pp. 2286--2303, May 2012.

\bibitem{DremeauHerzet12}
A.~Dr\'{a}meau, C.~Herzet, and L.~Daudet, ``Boltzmann machine and
mean-field
  approximation for structured sparse decompositions,'' \emph{IEEE Trans.
  Signal Processing}, vol.~60, no.~7, pp. 3425--3438, July 2012.

\bibitem{ZhangRao13}
Z.~Zhang and B.~D. Rao, ``Extension of {SBL} algorithms for the
recovery of
  block sparse signals with intra-block correlation,'' \emph{IEEE Trans. Signal
  Processing}, vol.~61, no.~8, pp. 2009--2015, Apr. 2013.

\bibitem{FangShen15}
J.~Fang, Y.~Shen, H.~Li, and P.~Wang, ``Pattern-coupled sparse
{B}ayesian
  learning for recovery of block-sparse signals,'' \emph{IEEE Trans. Signal
  Processing}, no.~2, pp. 360--372, Jan. 2015.

\bibitem{Rangan11}
S.~Rangan, ``Generalized approximate message passing for
estimation with random
  linear mixing,'' in \emph{Proc. IEEE Int. Symp. Inf. Theory (ISIT)}, Saint
  Petersburg, Russia, Aug. 2011.

\bibitem{DonohoMaleki10}
D.~L. Donoho, A.~Maleki, and A.~Montanari, ``Message passing
algorithms for
  compressed sensing: {I}. motivation and construction,'' in \emph{Proc. Inf.
  Theory Workshop}, Cairo, Egypt, Jan. 2010.

\bibitem{CandesTao05}
E.~Cand\'{e}s and T.~Tao, ``Decoding by linear programming,''
\emph{IEEE Trans.
  Information Theory}, no.~12, pp. 4203--4215, Dec. 2005.

\bibitem{VilaSchniter14}
J.~P. Vila and P.~Schniter, ``An empiricla-{B}ayes approach to
recovering
  linearly constrained non-negative sparse signals,'' \emph{IEEE Trans. Signal
  Processing}, vol.~62, no.~18, pp. 4689--4703, Sep 2014.

\bibitem{VilaSchniter13}
------, ``Expectation-maximization {G}aussian-mixture approximate message
  passing,'' \emph{IEEE Trans. Signal Processing}, no.~19, pp. 4658--4672, Oct.
  2013.

\bibitem{BergFriedlander08}
E.~van~den Berg and M.~P. Friedlander, ``Probing the {P}areto
frontier for
  basis pursuit solutions,'' \emph{SIAM J. Sci. Comput}, vol.~31, no.~2, pp.
  890--912, 2008.

\bibitem{WarnellReddy12}
G.~Warnell, D.~Reddy, and R.~Chellappa, ``Adaptive rate
compressive sensing for
  background subtraction,'' in \emph{IEEE International Conference on
  Acoustics, Speech and Signal Processing (ICASSP)}, Kyoto, Japan, March 25-30
  2012.

\bibitem{VilaSchniter11}
J.~P. Vila and P.~Schniter, ``Expectation-maximization
{B}ernoulli-{G}aussian
  approximate message passing,'' in \emph{45th Asilomar Conference on Signals,
  Symtems and computers}, Pacific Grove, CA, USA, November 6-9 2011.

\end{thebibliography}

\end{document}